\documentclass[10pt]{article}
\usepackage{soul}
\usepackage{multirow}
\usepackage{url,graphicx}
\usepackage{amsmath,amssymb}
\usepackage{algorithm}
\usepackage{wrapfig}
\usepackage{color}
\usepackage{rotating}
\usepackage{bm}
\usepackage{algorithmic}
\usepackage{subfig}

\usepackage[labelfont=bf,labelsep=period,justification=raggedright]{caption}

\makeatletter
\renewcommand{\@biblabel}[1]{\quad#1.}
\makeatother

\date{}

\begin{document}

\begin{flushleft}
{\Large
\textbf{Network-based Isoform Quantification with RNA-Seq Data for Cancer Transcriptome Analysis}
}
\\
Wei Zhang$^{1}$,
Jae-Woong Chang$^{2}$,
Lilong Lin$^{3}$,
Kay Minn$^{4}$,
Baolin Wu$^{5}$,
Jeremy Chien$^{4}$,
Jeongsik Yong$^{2}$,
Hui Zheng$^{3}$,
and Rui Kuang$^{1,\ast}$
\\
\bf{1} Department of Computer Science and Engineering, University of Minnesota Twin Cities, Minneapolis, Minnesota, United States of America
\\
\bf{2} Department of Biochemistry, Molecular Biology and Biophysics, University of Minnesota Twin Cities, Minneapolis, Minnesota, United States of America
\\
\bf{3} Guangzhou Institutes of Biomedicine and Health, Chinese Academy of Sciences, Guangzhou, Peoples Republic of China
\\
\bf{4} Department of Cancer Biology, University of Kansas Medical Center, Kansas City, Kansas, United States of America
\\
\bf{5} Division of Biostatistics, School of Public Health, University of Minnesota Twin Cities, Minneapolis, Minnesota, United States of America
\\
$\ast$ E-mail: kuang@cs.umn.edu
\end{flushleft}

\section*{Abstract}
High-throughput mRNA sequencing (RNA-Seq) is widely used for transcript quantification of gene isoforms. Since RNA-Seq data alone is often not sufficient to accurately identify the read origins from the isoforms for quantification, we propose to explore protein domain-domain interactions as prior knowledge for integrative analysis with RNA-Seq data. We introduce a Network-based method for RNA-Seq-based Transcript Quantification (Net-RSTQ) to integrate protein domain-domain interaction network with short read alignments for transcript abundance estimation. Based on our observation that the abundances of the neighboring isoforms by domain-domain interactions in the network are positively correlated, Net-RSTQ models the expression of the neighboring transcripts as Dirichlet priors on the likelihood of the observed read alignments against the transcripts in one gene. The transcript abundances of all the genes are then jointly estimated with alternating optimization of multiple EM problems.
In simulation Net-RSTQ effectively improved isoform transcript quantifications when isoform co-expressions correlate with their interactions. qRT-PCR results on 25 multi-isoform genes in a stem cell line, an ovarian cancer cell line, and a breast cancer cell line also showed that Net-RSTQ estimated more consistent isoform proportions with RNA-Seq data. In the experiments on the RNA-Seq data in The Cancer Genome Atlas (TCGA), the transcript abundances estimated by Net-RSTQ are more informative for patient sample classification of ovarian cancer, breast cancer and lung cancer. All experimental results collectively support that Net-RSTQ is a promising approach for isoform quantification. Net-RSTQ toolbox is available at \url{http://compbio.cs.umn.edu/Net-RSTQ/}.

\section*{Author Summary}
New sequencing technologies for transcriptome-wide profiling of RNAs have greatly promoted the interest in isoform-based functional characterizations of a cellular system. Elucidation of gene expressions at the isoform resolution could lead to new molecular mechanisms such as gene-regulations and alternative splicings, and potentially better molecular signals for phenotype predictions. However, it could be overly optimistic to derive the proportion of the isoforms of a gene solely based on short read alignments. Inherently, systematical sampling biases from RNA library preparation and ambiguity of read origins in overlapping isoforms pose a problem in reliability. The work in this paper exams the possibility of using protein domain-domain interactions as prior knowledge in isoform transcript quantification. We first made the observation that protein domain-domain interactions positively correlate with isoform co-expressions in TCGA data and then designed a probabilistic EM approach to integrate domain-domain interactions with short read alignments for estimation of isoform proportions. Validated by qRT-PCR experiments on three cell lines, simulations and classifications of TCGA patient samples in several cancer types, Net-RSTQ is proven a useful tool for isoform-based analysis in functional genomes and systems biology.

\section*{Introduction}
Application of next generation sequencing technologies to mRNA sequencing (RNA-Seq) is a widely used approach in transcriptome study \cite{NatureRNASeq1,NatureRNASeq2,PNAsSLIDE}. Compared with microarray technologies, RNA-Seq provides information for expression analysis at transcript level and avoids the limitations of cross-hybridization and restricted range of the measured expression levels. Thus, RNA-Seq is particularly useful for quantification of isoform transcript expressions and identification of novel isoforms. Accurate RNA-Seq-based transcript quantification is a crucial step in other downstream transcriptome analyses such as isoform function prediction in the pioneer work in \cite{ZhouUSC}, and differential gene expression analysis \cite{TaoDiff} or transcript expression analysis \cite{cuffdiff}. Detecting biomarkers from transcript quantifications by RNA-Seq is also a frequent common practice in biomedical research. However, transcript quantification is challenging since a variety of systematical sampling biases have been observed in RNA-Seq data as a result of library preparation protocols \cite{Bias1,LiRNASeq,Bias2,HuangRNASeq}. Moreover, in the aligned RNA-Seq short reads, most reads mapped to a gene are potentially originated by more than one transcript. The ambiguous mapping could result in hardly identifiable patterns of transcript variants \cite{HuangRNASeq,JiangRNASeq}.

\begin{figure}[H]
\centering
\begin{tabular}{c}
\vspace{-1mm}
{\scalebox{0.5}{\includegraphics*{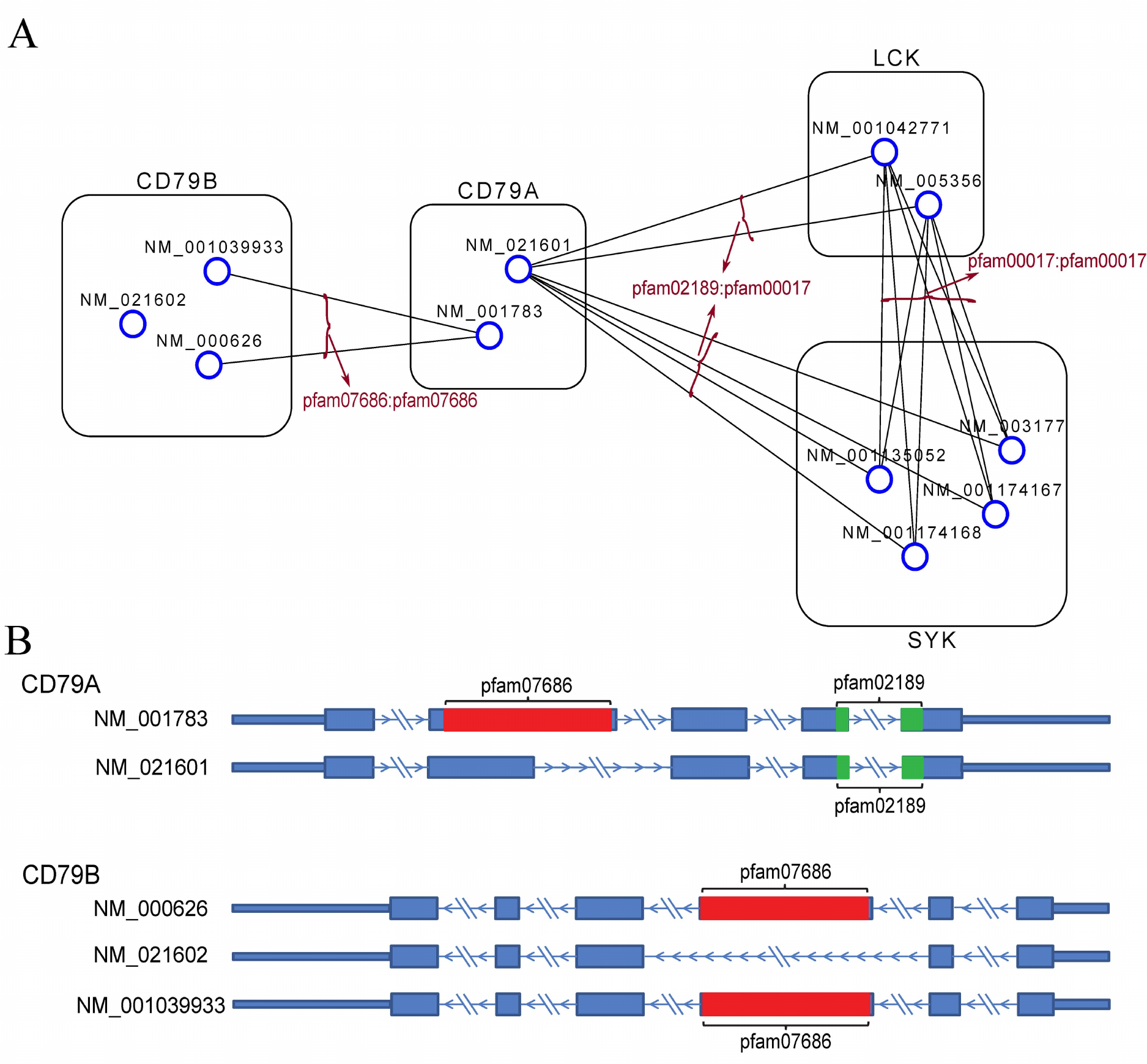}}}
\end{tabular}
\vspace{-2mm}
\caption{{\bf An isoform transcript network based on protein domain-domain interactions.} (A) The subnetwork shows the domain-domain interactions among transcripts from four human genes, CD79B, CD79A, LCK and SYK. In the network, the nodes represent isoform transcripts, which are further grouped and annotated by their gene name; and the edges represent domain-domain interactions between two transcripts. Each edge is also annotated by the interacting domains in the two transcripts. (B) RefSeq transcript annotations of CD79A and CD79B are shown with Pfam domain marked in color. The Pfam domains were detected with Pfam-Scan software. Note that no interaction is included between transcripts NM\_001039933 and NM\_000626 of gene CD79B without assuming self-interactions for modeling simplicity. For better visualization, only the interactions coincide with PPI are shown in the figure.
}\label{fig:intro}
\vspace{-1mm}
\end{figure}

A useful prior knowledge that has been largely ignored in RNA-Seq transcriptome quantification is the relation among the isoform transcripts by the interactions between their protein products. The protein products of different isoforms coded by the same gene may contain different domains interacting with the protein products of the transcripts in other genes. Previous studies suggested that alternative splicing events tend to insert or delete complete protein domains/functional motifs \cite{Functional} to mediate key linkages in protein interaction networks by removal of protein domain-domain interactions \cite{Functional2}. The work in \cite{ZhouUSC, isoform-coexp} also suggested unique patterns in isoform co-expressions. Thus, the abundance of an isoform transcript in a gene can significantly impact the quantification of the transcripts in other genes when their protein products interact with each other to accomplish a common function as illustrated by a real subnetwork in Fig \ref{fig:intro}, which is constructed based on domain-domain interaction databases \cite{iPfam, 3did} and Pfam \cite{Pfam}. Motivated by our observation that the protein products of highly co-expressed transcripts are more likely to interact with each other by protein domain-domain binding in four TCGA RNA-Seq datasets (see the section {\bf Results}), we constructed two human transcript interaction networks of different sizes based on protein domain-domain interactions to improve transcript quantification. Based on the constructed transcript network, we propose a network-based transcript quantification model called Net-RSTQ to explore domain-domain interaction information for estimating transcript abundance. In the Net-RSTQ model, Dirichlet prior representing prior information in the transcript interaction network is introduced into the likelihood function of observing the short read alignments. The new likelihood function of Net-RSTQ can be alternating-optimized over each gene with expectation maximization (EM). It is important to note that the Dirichlet prior from the neighboring isoforms play two possible roles. On one hand, for the isoforms in the same gene but with different interacting partners, the different prior information will help differentiate their expressions to reflect their different functional roles. On the other hand, for the isoforms in the same gene with the same interacting partners, the uniform prior assumes no difference in their functional roles and thus, promotes a smoother expression patterns across the isoforms. In both cases, the Dirichlet prior captures the functional variations/similarities across the isoforms in each gene as prior information for estimation of their abundance.

The paper is organized as following. In the section {\bf Materials and Methods}, we describe the procedure to construct protein domain-domain interaction networks, the mathematic description of the probabilistic model and the Net-RSTQ algorithm, qRT-PCR experiment design, and RNA-Seq data preparation. In the section {\bf Results}, we first demonstrate the correlation between protein domain-domain interactions and isoform transcript co-expressions across samples in four cancer RNA-Seq datasets from The Cancer Genome Atlas (TCGA) to justify using domain-domain interactions as prior knowledge. We then compared the predicted isoform proportions with qRT-PCR experiments on 25 multi-isoform genes in three cell lines, H9 stem cell line, OVCAR8 ovarian cancer cell line and MCF7 breast cancer cell line. Net-RSTQ was also applied to four cancer RNA-Seq datasets to quantify isoform expressions to classify patient samples by the survival or relapse outcomes. In addition, simulations were also performed to measure the statistical robustness of Net-RSTQ over randomized networks.

\section*{Materials and Methods}
In this section, we first describe the construction of the transcript interaction network and review the base probabilistic model for transcript quantification with RNA-Seq data. We then introduce the network-based transcript quantification model (Net-RSTQ) by applying the protein domain-domain interaction information as prior knowledge. The notations used in the equations are summarized in Table \ref{tab:notation}. At last, qRT-PCR experiment design and RNA-Seq data preparation are explained.
\begin{table}[h]
\small
\centering
\begin{tabular}{c|l}
\hline
{\bf Notation}&{\bf Description}\\
\hline
{$N$}&{total \# of genes}\\
{${\bm T}$}&{set of transcripts; $T_{ik}$ is the $k^{th}$ transcript of the $i^{th}$ gene; ${\bm T_{i}}$ denotes the transcripts of the $i^{th}$ gene} \\
{$l_{ik}$}&{length of transcript $T_{ik}$}\\
{${\bm r}$}&{set of reads; $r_{ij}$ is the $j^{th}$ read aligned to the $i^{th}$ gene; ${\bm r_{i}}$ is the read set aligned to the $i^{th}$ gene}\\
{$p_{ik}$}&{the probability of a read generated by transcript $T_{ik}$ in the $i^{th}$ gene}\\
{${\bm P_i}$}&{the probability of a read generated by transcript ${\bm T_{i}}$ in the $i^{th}$ gene, specifically, $[p_{i1},...,p_{i,|{\bm T_{i}}|}]$}\\
${\bm P}$ & {concatenate of all ${\bm P_i}$, specifically, $[{\bm P_1}, {\bm P_2}, ..., {\bm P_N}]$}\\
${\rho}_{ik}$ & {relative abundance of the transcript $T_{ik}$ in the $i^{th}$ gene}\\
{${\bm \pi}$}&{transcript expression; $\pi_{ik}$ is the expression of the $k^{th}$ transcript of the $i^{th}$ gene}\\
{${\phi}_{ik}$}&{average expressions (normalized) of transcript $T_{ik}$'s neighbors in the transcript network}\\
{${\bm \alpha}$}&{parameters of Dirichlet distribution; $\alpha_{ik}=\lambda\phi_{ik}+1$ is the parameter of the Dirichlet distribution of $p_{ik}$}\\
{$q_{ijk}$}&{read sampling probability, $q_{ijk}=\frac{1}{l_{ik}-l_r+1}$ if read $r_{ij}$ is aligned to transcript $T_{ik}$, otherwise $q_{ijk}=0$}\\
{${\bm S}$}&{binary matrix for transcript interaction network}\\
\hline
\end{tabular}
\caption{{\bf Notations}} \label{tab:notation}
\end{table}

\subsection*{Transcript network construction}\label{sec:network}
Two binary transcript networks were constructed by measuring the protein domain-domain interactions (DDI) between the domains in each pair of transcripts in four steps. First, the translated transcript sequences of all human genes were obtained from RefSeq \cite{RefSeq}. Second, Pfam-Scan was used to search Pfam databases for the matched Pfam domains on each transcript with 1e-5 e-value cutoff \cite{Pfam}. Note that only high quality, manually curated Pfam-A entries in the database were used in the search. Third, domain-domain interactions were obtained from several domain-domain interaction databases, and if any domain-domain interaction exists between a pair of transcripts, the two transcripts are connected in the transcript network. Specifically, 6634 interactions between 4346 Pfam domain families from two 3D structure-based DDI datasets (iPfam \cite{iPfam} and 3did \cite{3did}) inferred from the protein structures in Protein Data Bank (PDB) \cite{PDB} were used in the experiments.
Besides these highly confident structure-based DDIs, transcript interactions constructed from 2989 predicted high-confidence DDIs and 2537 predicted medium-confidence DDIs in DOMINE \cite{DOMINE} were also included if the transcript interaction agrees with protein-protein interactions (PPI) in HPRD \cite{HPRD}.

In the experiments, we focused on the transcripts from two cancer gene lists from the literature for better reliability in annotations. The first smaller transcript network consists of 11736 interactions constructed from the 3D structure-based DDIs and 421 interactions constructed from the predicted DDIs among the 898 transcripts in 397 genes from the first gene list \cite{COSMIC}. The second larger transcript network contains 711,516 interactions constructed from the 3D structure-based DDIs among 5599 transcripts in 2551 genes in a larger gene list \cite{SKCC}. Since inclusion of the predicted DDIs results in a much higher density in the large network, the large network does not include predicted DDIs to prevent too many potential false positive interactions. The characteristics of the two transcript networks are summarized in Table \ref{tab:newtworkInfo}. The density of the two networks are 3.02\% and 4.54\% respectively, which are in similar scale with the PPI network.  Both networks show high clustering coefficients, suggesting modularity of subnetworks.
Note that self-interactions (interactions between transcript(s) in the same gene) are not considered since Net-RSTQ only utilizes positive correlation between the expressions of neighboring transcripts in different genes. For simplicity, Net-RSTQ assumes that self-interactions will not change the transcript quantification of an individual gene in the model.

\begin{table}[H]
\centering
\tiny
\begin{tabular}{|c|c|c|c|c|c|c|c|}
\hline
&{\# of Gene}&{\# of Transcripts}&{\# of Interactions}&{Density}&{Diameter}&{Avg. \# of Neighbors}&{Avg. Cluster Coefficients}\\
\hline
{Small Network}&{397}&{898}&{12157}&{3.02\%}&{9}&{27.08}&{0.3578}\\
\hline
{Large Network}&{2551}&{5599}&{711516}&{4.54\%}&{9}&{254.16}&{0.5255}\\
\hline
\end{tabular}\hfill
\caption{{\bf Network characteristics.} } \label{tab:newtworkInfo}
\end{table}

In Fig \ref{fig:intro}(A) a subnetwork of the transcripts in gene CD79A and CD79B with their direct neighbors in the small transcript network is shown. The RefSeq transcript annotations of CD79A and CD79B are shown in Fig \ref{fig:intro}(B). In CD79A transcript NM\_001783 contains an extra domain pfam07686 while transcript NM\_021601 only contains a shorter hit pfam02189. Note pfam02189 also has the same hit in NM\_001783 with an e-value larger than 1e-5. In CD79B transcripts NM\_001039933 and NM\_000626 contain a domain pfam07686, which is removed in alternative splicing of NM\_021602. In the transcript subnetwork shown in Fig \ref{fig:intro}(A), the transcripts in CD79A or CD79B have different interaction partners in the network. In the transcripts in CD79A, the expression of NM\_021601 will correlate with the transcripts in LCK and SYK, and NM\_001783 will correlate with two transcripts in CD79B. The isoform transcripts in LCK and SYK show no different DDIs suggesting there is no functional variation by protein bindings and more similar expression patterns are potentially expected as prior knowledge.

\subsection*{Base model for transcript quantification} \label{sec:base}

We first consider the method proposed in \cite{FirstModel,RSEM2010} as the base model for quantification of the transcripts in a single gene. Let $\bm T_{i}$ denote the set of the transcripts in the $i$th gene and $T_{ik}$ be the $k$th transcript in $\bm T_i$. The probability of a read being generated by the transcripts in $\bm T_{i}$ is modeled by a categorical distribution specified by parameters $p_{ik}$, where $\sum_{k=1}^{|{\bm T_i}|}p_{ik} = 1$ and $0 \leq p_{ik} \leq 1$. For the set of the reads $\bm r_i$ aligned to gene $i$, we consider the likelihood of that each of the $|\bm r_i|$ short reads is sampled from one of the transcripts to which the read aligns. Specifically, for each read $r_{ij}$ aligned to transcript $T_{ik}$, the probability of obtaining $r_{ij}$ by sampling from $T_{ik}$, namely $Pr(r_{ij}|T_{ik})$ is $q_{ijk}=\frac{1}{l_{ik}-l_r+1}$ \cite{mgmr,LiRNASeq,RNASeqReview}, where $l_r$ is the length of the read. Assuming each read is independently sampled from one transcript, the uncommitted likelihood function  \cite{FirstModel}  to estimate the parameters ${\bm P_i}$ from the observed read alignments against gene $i$ is
\begin{eqnarray}	
    \mathcal{L}({\bm P_i};{\bm r_i}) = Pr({\bm r_i}|{\bm P_i}) &=& \prod_{j=1}^{|\bm r_i|}Pr(r_{ij}|{\bm P_i}) = \prod_{j=1}^{|\bm r_i|}\sum_{k=1}^{|\bm T_i|}Pr(T_{ik}|{\bm P_i})Pr(r_{ij}|T_{ik}) = \prod_{j=1}^{|\bm r_i|}\sum_{k=1}^{|\bm T_i|}p_{ik}q_{ijk}. \label{eqn:likelihood1}
\end{eqnarray}
This likelihood function is concave but it may contain plateau in the likelihood surface. Therefore, Expectation Maximization (EM) is then applied to obtain the optimal ${\bm P_i}$. In the EM algorithm, the expectation of read assignments to transcripts were estimated in the E-step and the likelihood function with the expected assignments can be maximized in the M-step to estimate  ${\bm P_i}$. The relative abundance of the transcript $T_{ik}$ in gene $i$, ${\rho}_{ik}$, can be derived from

\begin{eqnarray}
    {\rho}_{ik} = \frac{\frac{p_{ik}}{l_{ik}}}{\sum_{k=1}^{|\bm T_i|}\frac{p_{ik}}{l_{ik}}},
    \label{eqn:relativeAbundance}
\end{eqnarray}
and the transcript expressions in gene $i$, ${\pi}_{ik}$, can be calculated by
\begin{eqnarray}
    {\pi}_{ik}=\frac{|r_{i}|p_{ik}}{l_{ik}}.
    \label{eqn:expression}
\end{eqnarray}
The base model is applied independently to each individual gene and no relation among the transcripts is considered.

\subsection*{Network-based transcript quantification model}
\begin{figure}[H]
\centering
\begin{tabular}{c}
{\scalebox{1}{\includegraphics*{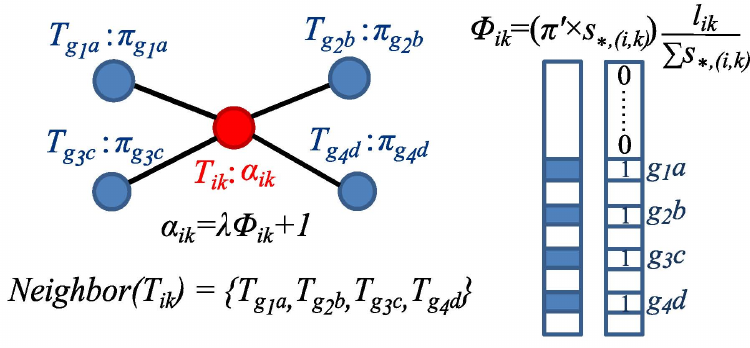}}}
\end{tabular}
\caption{{\bf Transcript interaction neighborhood.} In this toy example, transcript $T_{ik}$ has four neighbor transcripts $\{T_{g_1a}, T_{g_2b}, T_{g_3c}, T_{g_4d}\}$, which are transcripts from $g_1$, $g_2$, $g_3$ and $g_4$, respectively. The neighborhood expression $\phi_{ik}$ of $T_{ik}$ is then calculated as the average of its neighbor transcripts' expressions and further normalized by transcript length, represented as the vector product between $\bm \pi$ and $\bm S_{*,(i,k)}$ normalized by the number of neighbors $\sum{\bm S}_{*,(i,k)}$ and the transcript length $l_{ik}$ in the figure.}\label{fig:sumNeighbor}
\end{figure}

In the Net-RSTQ model, the transcript interaction network ${\bm S}$ based on protein domain-domain interactions is introduced to calculate a prior distribution for estimating ${\bm P}$ jointly across all the genes and all the transcripts. The model assumes that the prior distribution of $\bm P_i$ is a Dirichlet distribution specified by parameters $\bm \alpha_{i}$ and each $\alpha_{ik}$ is proportional to the read count by average expression of the transcript $T_{ik}$'s neighbors in the transcript network ${\bm S}$. The prior read count ${\phi}_{ik}$ is defined as follows,

\begin{eqnarray}
    {\phi}_{ik}={l_{ik}}({\bm \pi^{'}}\frac{{\bm S}_{*,(i,k)}}{\sum({\bm S}_{*,(i,k)})}),
    \label{eqn:normalizedPhi}
\end{eqnarray}
where ${\bm S}_{*,(i,k)}$ is a binary vector represents the neighborhood of transcript $T_{ik}$ in transcript network ${\bm S}$ and ${\sum({\bm S}_{*,(i,k)})}$ is the size of the neighborhood. The calculation of each ${\phi}_{ik}$ is illustrated in Fig \ref{fig:sumNeighbor}. The Dirichlet parameter ${\bm {\alpha}_i}$ is defined as a function of ${\phi}_{ik}$ as
\begin{eqnarray}
\alpha_{ik}=\lambda\phi_{ik}+1, \label{eqn:alpha}
\end{eqnarray}
where $\lambda > 0$ is a tuning parameter balancing the belief between the prior-read count and the aligned-read count.

To obtain the optimal $\bm P$ jointly for all genes, we introduce a pseudo-likelihood model to estimate $\bm P$ iteratively in each iteration. Assuming uniform $Pr({\bm r_i})$, the pseudo-likelihood function is defined as,
\begin{eqnarray} \label{eqn:glh}
    \mathcal{L}({\bm P},{\bm \alpha};{\bm r}) =  \prod_{i=1}^{N} \mathcal{L}({\bm P_i},{\bm \alpha_i};{\bm r_i}) = \prod_{i=1}^{N}  \frac{Pr({\bm P_i}|{\bm \alpha_i})Pr({\bm r_i}|{\bm P_i})}{Pr({\bm r_i})} \propto \prod_{i=1}^{N} Pr({\bm P_i}|{\bm \alpha_i})Pr({\bm r_i}|{\bm P_i}).
\end{eqnarray}
Note that the pseudo-likelihood model relies on the independence assumption among the likelihood functions of each individual gene when the $\bm \alpha$  parameters of the Dirichlet priors are pre-computed. Thus, the model simply takes the product of the likelihood function from each gene. Each prior distribution $Pr({\bm P_i}|{\bm \alpha_i})$ follows the Dirichlet distribution,

\begin{eqnarray} \label{eqn:DPrior}
    Pr({\bm P_i}|{\bm \alpha_i}) = C({\bm \alpha_i})\prod_{k=1}^{|\bm T_i|}{p_{ik}}^{\alpha_{ik}-1},
   \textrm{where } C({\bm \alpha_i}) = \frac{\Gamma(\sum_{k}\alpha_{ik})}{\prod_{k}\Gamma(\alpha_{ik})}.
\end{eqnarray}

Integrating equations {\eqref{eqn:likelihood1} and \eqref{eqn:DPrior}}, the pseudo-likelihood function in equation \eqref{eqn:glh} can be rewritten with Dirichlet prior  as 

\begin{eqnarray}
    \mathcal{L}({\bm P};{\bm r})&=&\prod_{i=1}^{N}\left[C({\bm \alpha_i})\prod_{k=1}^{|\bm T_i|}{p_{ik}}^{\alpha_{ik}-1}\right]\left[\prod_{j=1}^{|\bm r_i|}\sum_{k=1}^{|\bm T_i|}p_{ik}q_{ijk}\right] \nonumber\\
        &=&\prod_{i=1}^{N} \left[C(\lambda{\bm \phi_{i}}+1)\prod_{k=1}^{|\bm T_i|}{p_{ik}}^{{\lambda}{\phi}_{ik}}\right]\left[\prod_{j=1}^{|\bm r_i|}\sum_{k=1}^{|\bm T_i|}p_{ik}q_{ijk}\right].
    \label{eqn:likelihood2}
\end{eqnarray}

In the pseudo-likelihood function in equation {\eqref{eqn:likelihood2}}, the only hyper-parameter $\lambda$ balances the proportion between the Dirichlet priors and the observed read counts of each transcript. The larger the $\lambda$, the more belief put on the priors.

\subsection*{The Net-RSTQ algorithm}
The Net-RSTQ algorithm optimizes equation {\eqref{eqn:likelihood2}} by dividing the optimization into sub-optimization problems of sequentially estimating each ${\bm P_i}$. Specifically, we fix all ${\bm P}_c$, $c\neq i$, and thus ${\bm \phi}_{i}$ when estimating ${\bm P_i}$ with EM in each iteration and repeat the process multiple rounds throughout all the genes. In each step, the neighborhood expression ${\bm \phi}$ is recomputed with new ${\bm P_i}$ for computing the quantification of the next gene. For each sub-optimization problem, we estimate ${\bm P_i}$ with a fixed ${\bm \phi}$, the part of the likelihood function in equation {\eqref{eqn:likelihood2}} involved with the current variables $\bm P_i$ is
\begin{eqnarray}
    \mathcal{\bar L}({\bm P_i};{\bm r_i})&=&\left[\prod_{g \in {\bm {nb}(i)}}C(\lambda{\bm \phi_{g}}+1)\prod_{k=1}^{|\bm T_g|}{p_{gk}}^{{\lambda}{\phi}_{gk}}\right]\left[C(\lambda{\bm \phi_{i}}+1)\prod_{k=1}^{|\bm T_i|}{p_{ik}}^{{\lambda}{\phi}_{ik}}\right]\left[\prod_{j=1}^{|\bm r_i|}\sum_{k=1}^{|\bm T_i|}p_{ik}q_{ijk}\right],
    \label{eqn:sublikelihoodA}
\end{eqnarray}
where $\bm {nb}(i)$ is the set of the genes containing transcripts that are neighbors of the transcripts in gene $i$ in the transcript network. Equation {\eqref{eqn:sublikelihoodA}} consists of three terms separated by the braces. The second and the third terms are the Dirichlet prior and the likelihood of the observed counts in the data for gene $i$. The first term is the Dirichlet priors of the neighbor transcripts of each $T_{ik}$. These prior probabilities are involved since ${\bm \phi_{g}}$ are functions of the current variable $\bm P_{i}$ (equations \eqref{eqn:expression}-\eqref{eqn:alpha}). Equation {\eqref{eqn:sublikelihoodA}} cannot be easily solved with standard techniques. We adopt a heuristic approach to only take steps that will increase the whole pseudo-likelihood function in equation {\eqref{eqn:likelihood2}}. The Net-RSTQ algorithm is outlined below
\begin{algorithm}[H]
    \small
	\caption{\ensuremath{\text{\sc Net-RSTQ}}}
	\label{alg:Net-RSTQ}
	\begin{algorithmic}[1]
    \STATE {\bf Initialization}: random initialization or base EM (equation \eqref{eqn:likelihood1}) estimation of ${\bm P}^{(0)}$
    \FOR{round $t=1,\ldots$}
        \STATE{${\bm P}^{(t)}={\bm P}^{(t-1)}$}
        \FOR{gene $i=1,\ldots,N$}
            \STATE{compute ${\bm \phi_i}$ based on ${\bm P}^{(t)}$ with equations {\eqref{eqn:expression}} and {\eqref{eqn:normalizedPhi}}}
            \STATE{estimate ${\bm P_i}$ with EM algorithm (see next section)}
            \IF{$\mathcal{\bar L}(\bm P_i) > \mathcal{\bar L}(\bm P_i ^ {(t)})$}
                \STATE{${\bm P_i}^{(t)}={\bm P_i}$}
            \ENDIF
        \ENDFOR
         \IF{$max(abs({\bm P}^{(t)}-{\bm P}^{(t-1)}))<$1e-6}
                \STATE{\textbf{break}}
         \ENDIF
    \ENDFOR
    \RETURN ${\bm P}$
    \end{algorithmic}
\end{algorithm}
In the algorithm, the outer for-loop between line 2-14 performs multiple passes of updating ${\bm P}$. The inner for-loop between line 4-10 scans through each gene to update each ${\bm P_i}$. Line 7 checks the the difference in the likelihood $\mathcal{\bar L}$ of gene $i$ before and after the estimated ${\bm P_i}$ is applied. The newly estimated ${\bm P_i}$ is kept in line 8 only if the likelihood $\mathcal{\bar L}$ in equation {\eqref{eqn:sublikelihoodA}} is higher.
The convergence of $\bm P$ is checked at line 11. In each sub-optimization problem, EM algorithm (described in the next section) is applied to estimate ${\bm P_i}$.
After convergence, the transcripts expression $\bm \pi$ can be learned by equation \eqref{eqn:expression} with the optimal $\bm P$.

\subsection*{Estimating ${\bm P_i}$ given ${\bm \phi_i}$}\label{sec:EM}
In line 6 of Algorithm 1, we maximize the likelihood function of the sub-optimization problem in equation {\eqref{eqn:sublikelihoodA}} to learn $\bm P_{i}$ as
\begin{eqnarray}
    \mathcal{L}({\bm P_i};{\bm r_i})=\left[C(\lambda{\bm \phi_{i}}+1)\prod_{k=1}^{|\bm T_i|}{p_{ik}}^{{\lambda}{\phi}_{ik}}\right]\left[\prod_{j=1}^{|\bm r_i|}\sum_{k=1}^{|\bm T_i|}p_{ik}q_{ijk}\right].
    \label{eqn:sublikelihoodA2}
\end{eqnarray}
Note that equation \eqref{eqn:sublikelihoodA2} is the part of equation \eqref{eqn:sublikelihoodA} without the Dirichlet priors of the neighboring genes. In line 7 of Algorithm 1, the ignored Dirichlet priors are combined with the likelihood in equation \eqref{eqn:sublikelihoodA2}, when $\mathcal{\bar L}(\bm{P_i})$ is computed, to evaluate the whole likelihood in equation \eqref{eqn:sublikelihoodA}. The likelihood function in equation \eqref{eqn:sublikelihoodA2} is defined on a categorical variable with Dirichlet prior, which can be solved with EM algorithm. Following EM formulation in \cite{mgmr}, the expectation $a_{ijk}$, a soft assignment of read $j$ to transcript $k$ in gene $i$, is first estimated in the expectation step and $\bm P_{i}$ is then learned in the maximization step. When ${\bm \phi_i}$ is given, by taking log of equation {\eqref{eqn:sublikelihoodA2}} we can write the EM steps to find ${\bm P_i}$ below.\\\textbf{E step:}

Letting $\textit{\textbf{Match}}$ signify a matching between reads and transcripts, and $\textit{Match(j)}$ be the transcript from which read $j$ originates, we get:
\begin{eqnarray}
    \log[\mathcal{L}({\bm P_i};{\bm r_i},{\bm {Match}})] &=& {\log}C(\lambda{\bm \phi_{i}}+1)+\sum_{k=1}^{|\bm T_i|}{\lambda}{\phi}_{ik}\log(p_{ik})+ \sum_{j=1}^{|\bm r_i|}\log(p_{iMatch(j)}q_{ijMatch(j)}),
\end{eqnarray}
which leads to
\begin{eqnarray}
    \mathcal{Q}({\bm P_i}|{\bm P}_{i}^{(it)}) &=& E_{{\bm {Match}}|{\bm r_i},{\bm P}_{i}^{(it)}}[\log(\mathcal{L}({\bm P_i};{\bm r_i}))] \nonumber\\
    &=& {\log}C(\lambda{\bm \phi_{i}}+1)+\sum_{k=1}^{|\bm T_i|}{\lambda}{\phi}_{ik}\log(p_{ik}^{(it)}) +\sum_{j=1}^{|\bm r_i|}\sum_{k=1}^{|\bm T_i|}({\log}p_{ik}^{(it)}+{\log}q_{ijk})*\frac{p_{ik}^{(it)}q_{ijk}}{\sum_{k=1}^{|\bm T_i|}p_{ik}^{(it)}q_{ijk}}\nonumber\\
    &=& {\log}C(\lambda{\bm \phi_{i}}+1)+\sum_{k=1}^{|\bm T_i|}{\lambda}{\phi}_{ik}\log(p_{ik}^{(it)})+ \sum_{j=1}^{|\bm r_i|}\sum_{k=1}^{|\bm T_i|}a_{ijk}\log(p_{ik}^{(it)}) + \sum_{j=1}^{|\bm r_i|}\sum_{k=1}^{|\bm T_i|}a_{ijk}\log(q_{ijk})
\end{eqnarray}
where $it$ is the ${it}^{th}$ iteration in EM and
\begin{eqnarray}
    a_{ijk} = \frac{p_{ik}^{(it)}q_{ijk}}{\sum_{k=1}^{|\bm T_i|}p_{ik}^{(it)}q_{ijk}}.
    \label{eqn:Estep}
\end{eqnarray}
\textbf{M step:}

Given that $q_{ijk}$ and ${\bm \phi_i}$ are known, the above reduces to maximizing
\begin{eqnarray}
    {\bm P}_{i}^{(it+1)} = \arg\max_{\substack{{\bm P_i}}}\left[\sum_{k=1}^{|\bm T_i|}\lambda{\phi}_{ik}log(p_{ik}) + \sum_{j=1}^{|\bm r_i|}\sum_{k=1}^{|\bm T_i|}a_{ijk}log(p_{ik})\right].
    \label{eqn:Mstep}
\end{eqnarray}
Using Lagrange multipliers and differentiating, equation {\eqref{eqn:Mstep}} is maximized when
\begin{eqnarray}
    p_{ik}^{(it+1)} = \frac{\lambda{\phi}_{ik}+\sum_{j=1}^{|\bm r_i|}a_{ijk}}{\sum_{k=1}^{|\bm T_i|}(\lambda{\phi}_{ik}+\sum_{j=1}^{|\bm r_i|}a_{ijk})}.
    \label{eqn:Mstep2}
\end{eqnarray}
After EM algorithm converges, we update $\bm P$ with the newly estimated ${\bm P_i}$ only if the update leads to increase of equation {\eqref{eqn:sublikelihoodA}}. It can be seen from equation \eqref{eqn:Mstep2} that the role of $\lambda$ is a parameter controlling the balance between the prior-read count and the aligned-read count. To see that, recall $\phi_{ik}$ is the prior-read count of transcript $T_{ik}$ by the average expression of its neighbors (equation \eqref{eqn:normalizedPhi}) and $\sum_{j=1}^{|\bm r_i|}a_{ijk}$ is the expected aligned-read count of transcript $T_{ik}$. $\lambda$ directly balances the contributions from the two terms. Therefore, a reasonable choice of $\lambda$ should apply to RNA-Seq data with similar level of noise or bias in general.

\subsection*{qRT-PCR experiment design}\label{sec:qPCR}

Three qRT-PCR experiments are designed to measure the isoform proportions of 25 multi-isoform genes in three cell lines, H9 stem cell line, OVCAR8 ovarian cancer cell line and MCF7 breast cancer cell line. The cell lines were selected based on the available of both RNA-Seq data and cell culture in our labs. The qRT-PCR experiments focused on the gene with most different quantification results reported by Net-RSTQ and other compared methods. Due to the limitations in time and cost of running qRT-PCR experiments, only the 25 genes in the three cell lines were tested with all the results reported in the experiments. Quantitation of the real-time PCR results was done on the data from H9 human embryonic stem cells to obtain the absolute expressions for comparing more than two transcripts and comparative Ct method was done on the data from OVCAR8 ovarian cancer cells and MCF7 breast cancer cells to obtain the ratio between a pair of transcripts.

\paragraph*{H9 Stem cell line:}
Total RNA was extracted from human embryonic stem (ES) H9 cells by using TRIzol (Invitrogen). To repeat the experiments of triplicate three times, 5${\mu}$g RNA was used to synthesize complementary DNA with ReverTra Ace (Toyobo) and oligo-dT (Takara) according to the manufacturer's instructions. Transcript levels of genes were determined by using Premix Ex Taq (Takara) and analysed with a CFX-96 Real Time system (Bio-Rad). The templates for different transcripts were generated with PCR by using the template primers in S1 Table in {\bf Supplementary}. After isolation and purification, the templates were used to generate the standard curves with qRT-PCR by using the qRT-PCR primers for different transcripts. The generated standard curves have coefficient of determination (R2) over 0.999. The qRT-PCR primers were then applied to determine the expression levels of different transcripts in H9 ES cells by calculating with the standard curves. The expressions were carried out in three independent replications and the standard deviations were provided after the average.

\paragraph*{Ovarian cancer cell line:}
1$\mu$g of total RNAs were isolated from untreated OVCAR8 cells using Trizol (Invitrogen). RNA was reverse-transcribed using Superscript II reverse transcriptase (Invitrogen) according to manufacture protocol. Real-time PCR was performed on CFX384 Real-time system (Bio-Rad) with FastStart SYBR Green Master (Roche) with the primer sets in S2 Table in {\bf Supplementary}. PCR conditions are 10 min at 95$^{\circ}$C and 40 cycles of 95$^{\circ}$C for 45 sec and 60$^{\circ}$C for 45 sec. Quantitation of the real-time PCR results was done using comparative Ct method. Two replicates of qRT-PCR were performed using total RNAs isolated.

\paragraph*{Breast cancer cell line:}
0.5${\mu}$g of total RNAs purified from MCF7 cells was used for oligo d(T$)_{20}$-primed reverse transcription (Superscript III; Life Technologies). SYBR Green was used to detect and quantitate PCR products in real-time reactions with the primer sets in S3 Table in {\bf Supplementary}. PCR conditions for qRT-PCR analysis are 2 min 94$^{\circ}$C and 40 cycles of 94$^{\circ}$C for 30 sec,  60$^{\circ}$C for 20 sec and 72$^{\circ}$C for 30 sec. Quantitation of the real-time PCR results was done using comparative Ct method. GAPDH mRNA was used as a normalization control for quantitation. Three replicates of qRT-PCR were performed using total RNAs isolated.

\subsection*{RNA-Seq data preparation}
Three cell line RNA-Seq datasets were used for evaluating the accuracy of transcript quantification by comparison with qRT-PCR results. The first dataset is the H9 embryonic stem cell line data from \cite{H9Data}, downloaded from SRA. The second dataset is an in-house dataset from the ovarian cancer cell line OVCAR8 prepared at University of Kansas Medical Center. The third dataset is the MCF7 breast cancer cell line data from \cite{MCF7Data}, downloaded from SRA. There are 23,397,325 single-end 34bp reads in the stem cell line dataset, 19,892,473 paired-end 100bp reads in the OVCAR8, and 21,855,632 paired-end 76bp reads in the MCF7 mapped to the human hg19 reference genome by TopHat2.0.9 \cite{Tophat2} with up to 2 mismatches allowed. Exon coverages and read counts of exon-exon junctions were generated by SAMtools \cite{Samtools} to be utilized with Net-RSTQ and base EM (equation \eqref{eqn:likelihood1}). Cufflinks \cite{cufflinks} directly infers transcript expressions based on the alignment by TopHat with the min isoform fraction set to 0 for better sensitivity.

TCGA RNA-Seq datasets of Ovarian serous cystadenocarcinoma (OV), Breast invasive carcinoma (BRCA), Lung adenocarcinoma (LUAD) and Lung squamous cell carcinoma (LUSC) were analyzed for patient outcome prediction with transcript expressions estimated by Net-RSTQ, base EM (equation \eqref{eqn:likelihood1}), RSEM \cite{RSEM} and Cufflinks  \cite{cufflinks}. Both the gene expression and transcript expression data reported by RSEM \cite{RSEM} in TCGA (level 3 data) were utilized as two baselines for cancer outcome prediction. The raw RNA-Seq fastq files (level 1 data) were downloaded from Cancer Genomics Hub (CGHub) and processed by TopHat for use with Net-RSTQ, base EM and Cufflinks. The patient samples in each dataset were classified into cases and controls based on the survival and relapse outcomes as shown in Table \ref{tab:dataset}. The command lines for preparing the data with RSEM and Cufflinks are available in the S3 Text in \textbf{Supplementary}.
\begin{table}
\centering
\begin{tabular}{|l|l|l|}
\hline
{\bf Cancer Type}&{\bf Event}&{\bf \# of Patients by years}\\
\hline
{Ovarian serous cystadenocarcinoma(OV)}&{Survival}&{ 76(\textless 3 ys) {\bf vs} 62(\textgreater 4 ys)}\\
&{Relapse}&{ 79(\textless 1.5 ys) {\bf vs} 68(\textgreater 2 ys)}\\
\hline
{Breast invasive carcinoma(BRCA)}&{Survival}&{66(\textless 5 ys) {\bf vs} 57(\textgreater 8 ys)}\\
&{Relapse}&{42(\textless 5 ys) {\bf vs} 38(\textgreater 8 ys)}\\
\hline
{Lung adenocarcinoma(LUAD)}&{Survival}&{47(\textless 2 ys) {\bf vs} 56(\textgreater 3 ys)}\\
\hline
{Lung squamous cell carcinoma(LUSC)}&{Survival}&{67(\textless 2 ys){\bf vs} 77 (\textgreater 3 ys)}\\
\hline
\end{tabular}\hfill
\caption{{\bf Summary of patient samples in TCGA datasets.} The samples are classified by cutoffs on survival and relapse time based on the available clinical information in each dataset.} \label{tab:dataset}
\end{table}

\section*{Results}
There are six major results in this section, 1) isoform co-expression analysis on TCGA data to show the correlation with protein domain-domain interactions; 2) overlapping the DDIs and KEGG pathways to understand the transcript networks; 3) simulations for model validation and statistical analysis; 4) qRT-PCR experiments to measure the performance of transcript quantification; 5) cancer outcome prediction on TCGA data to measure the quality of transcript quantification as molecular markers; and 6) running time of Net-RSTQ.

Net-RSTQ was compared with base EM (the base model in equation \eqref{eqn:likelihood1}), Cufflinks  \cite{cufflinks} and RSEM (isoform expression or gene expression) \cite{RSEM}. The accuracy of transcript quantification was directly measured on the simulated data with ground-truth expressions and qRT-PCR data from the three cell lines. Cancer outcome prediction on four TCGA cancer datasets evaluates the potential of using isoform expressions as predictive biomarkers in clinical settings. Statistical assessment was also performed on randomized transcript networks to evaluate the significance of the results.

\subsection*{Isoform co-expressions correlate with protein domain-domain interactions} \label{sec:co-exp}
\begin{figure}[!]
\centering
{\scalebox{0.8}{\includegraphics*{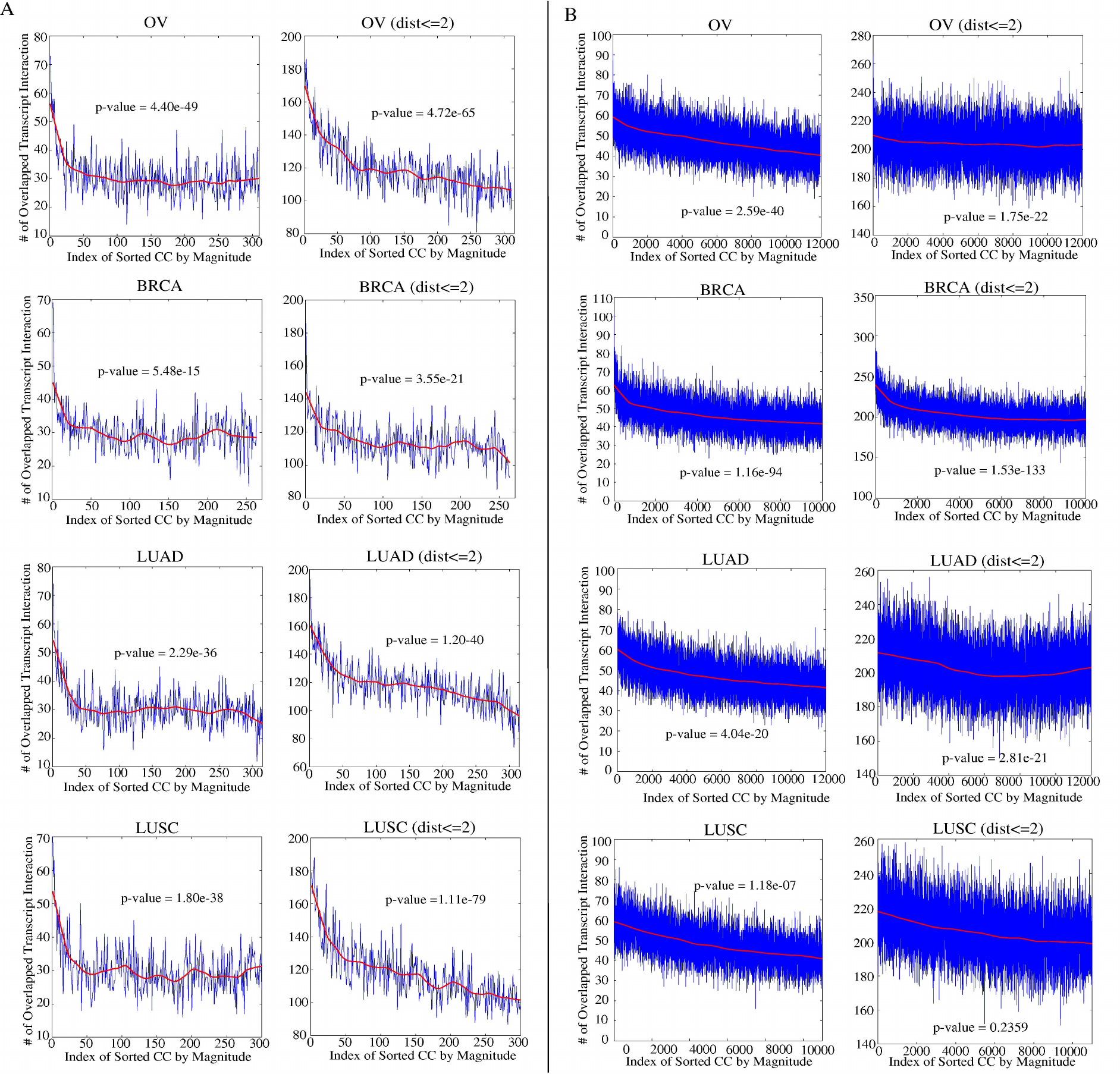}}}
\caption{{\bf Correlation between transcript co-expression and protein domain-domain interaction in TCGA datasets.} The correlation coefficients between transcript expressions across all patient samples are first calculated in each dataset for each pair of transcripts by Cufflinks. The correlation coefficients are then sorted from largest to smallest and grouped into bins of size 1000 each. The x-axis is the index of the bins with lower index indicating larger correlation coefficients. The y-axis is the number of the pairs among the 1000 pairs of transcripts in each bin coincide with protein domain-domain interaction between the transcript pair. The red line is the smooth plot by fitting local linear regression method with weighted linear least squares (LOWESS) to the curves. $p$-value is reported by chi-square test. (A) Co-expressions are calculated based on the small gene list. (B) Co-expressions are calculated based on the large gene list. In both (A) and (B), the left column shows the plots based on the connected transcript pairs in the transcript network and the right column shows the plots based on the transcript pairs with distance up to 2 in the network.}\label{fig:coexpression}
\end{figure}

To investigate the correlation between protein domain-domain interactions and isofrom transcript co-expressions, we calculated the number of transcript pairs that are both nearby (being neighbors or having a distance up to 2) in the transcript network and highly co-expressed in the TCGA samples. The transcript co-expressions were calculated by Pearson's correlation coefficients of each pair of transcripts across all the samples in each dataset with the isoform transcript quantification by Cufflinks. The transcript pairs were then sorted by the correlation coefficients from the largest to the smallest and grouped into bins of size 1000. The number of transcript pairs that are nearby in the transcript networks out of 1000 pairs are calculated within each bin and plotted in Fig \ref{fig:coexpression}(A) and Fig \ref{fig:coexpression}(B) for the two cancer gene lists, respectively. In both Fig \ref{fig:coexpression}(A) and Fig \ref{fig:coexpression}(B), the left column shows the plots of the number of pairs that are neighbors in the transcript network, and the right column shows the plots of the number of transcript pairs with a distance up to 2 in the transcript network, among the 1000 pairs in each bin.
In all the plots, similar trends are observed in all the four cancer datasets: there are more interacting isoform pairs in the bins with higher co-expressions. For example, among the 1000 transcript pairs with the highest correlation coefficients, there are 73 interactions in the transcript network in OV dataset and thus, 73 interactions (y-axis) for bin index 1 (x-axis) is plotted in the left column of Fig \ref{fig:coexpression}(A). In all the plots, there is a clear pattern that the numbers of matched nearby transcripts in the transcript network among the 1000 pairs in the first few bins are higher than the expected average of 30 in the small network of density 3.02\%, 114 in the small network of density 11.41\% (with distance up to 2), 45 in the larger network of density 4.54\%, and 203 in the larger network of density 20.33\% (with distance up to 2). Moreover, the 2-step walk clearly promoted the number of overlaps with the pairs of higher co-expressions in the small network. For example, the significant overlap is extended from the first 25 bins to approximately the first 50 bins or more in the four datasets. The observation suggests that higher co-expressions exist not only in the direct neighbors in the transcript network but also the nearby nodes by a small distance. By exploring the network structure with prior information through neighbors by many steps in iterations, Net-RSTQ model is expected to propagate the expression values from each transcript to its nearby nodes in the network to capture the co-expressions. Note that considering the neighboring pairs with distance up to 2 in the larger network will result in a graph of density 20.33\%, which is likely to contain too many false relations by the two-step walk. Thus, the plots of the larger network of distance-2 pairs are only included for the completeness of the analysis.

The canonical 2x2 chi-square test was also applied to compare the number of the domain-domain interactions in the first 10,000 transcript pairs (first 10 bins) with the number in the rest of the pairs. In all the four datasets in both Fig \ref{fig:coexpression}(A) and Fig \ref{fig:coexpression}(B) with one exception in the LUSC dataset on the large network of distance-2 relation, there is a significant difference that the highly co-expressed transcripts are more likely to interact with each other in the transcript network, confirmed by the significant $p$-values. As explained previously, the exception is likely due to the large number of false-positive pairs in the dense network. The observation further support the hypothesis that protein domain-domain interactions correlate transcript co-expressions reported in previous studies \cite{Functional,Functional2}.

To further understand the specificity of the domain-domain interactions in the highly co-expressed transcripts, we calculated the number of domain-domain pairs that construct the DDIs in the top 10,000 co-expressed transcript pairs. The statistics suggest high diversity of the type of DDIs. For example, there are 547 interacting transcript pairs among the 201 out of 898 transcripts in the top 10,000 co-expressed transcript pairs in OV dataset for small network. The 547 interacting transcript pairs represent 770 different domain-domain interactions (There might be more than one DDIs between a pair of transcripts). There are 739 interacting transcript pairs among the 538 out of 5599 transcripts in the top 10,000 co-expressed transcript pairs in OV dataset for large network. The 739 interacting transcript pairs represent 1277 different domain-domain interactions. The statistics suggest that the correlation between protein domain-domain interactions and transcript co-expressions is not a bias due to a few highly spurious DDIs. It is a general correlation in many different DDIs and co-expressed transcripts. Very similar statistics were observed in all the datasets and both networks.

To further demonstrate the co-expression relations in the transcript network, two examples are shown in S1 Fig. In S1(A) Fig, WHSC1L1 contains two isoforms connected with different interactions in the transcript network. Isoform NM\_017778 interacts with 12 transcripts with average correlation coefficients 0.22 and the other isoform NM\_023034 interacts with 13 more transcripts with average correlation coefficients 0.30 compared with the average correlation coefficient 0.188 against the other unconnected isoforms across the samples in the OV dataset. In S1(B) Fig, gene BRD4 contains two isoforms both of which are connected with the same 14 neighbors in the network. The average correlation coefficients between these two isoforms and the 14 neighboring isoforms are both above 0.26 compared with the average correlation coefficient less than 0.15 against the other unconnected isoforms across the samples on the BRCA dataset. In both examples, we observed high degree of agreement between co-expressions and DDIs.

\subsection*{Protein domain-domain interactions enrich KEGG pathways}
To further understand the transcript networks, we overlapped the DDIs between genes in the two networks with the 294 human KEGG pathways \cite{KEGG2000}. Among the 397 genes in the small network, 10.97\%(17284) of the pairs are co-members in at least one KEGG pathway. The 10.97\% KEGG co-member pairs covers 42.70\%(2122) of the DDIs among the genes while the other 89.03\%(140352) non-co-member pairs covers 57.30\%(2748) of the DDIs. By these numbers, there is about 6-fold enrichment of DDIs in the KEGG co-member genes in the small network.
Among the 2551 genes in the large network, the 5.15\%(335372) KEGG co-member pairs covers 12.45\%(40812) of the DDIs among genes while the other 94.85\%(6172229) non-co-member pairs covers 87.55\%(287090) of the DDIs. By these numbers, there is about 2.6-fold enrichment of DDIs in the KEGG co-member genes in the large network.
We also list the KEGG pathways that are highly enriched with DDIs in the large network in S4 Table. Specifically, we consider the subnetwork of genes that are members of one KEGG pathway and calculated the density of DDIs in the subnetwork to compare to the overall density of 5.04\% in the whole network. Interestingly, most of the enriched pathways are signaling pathways and disease pathways with very high DDI densities.

\subsection*{Net-RSTQ captures network prior in simulations}
In the simulations, we applied flux-simulator \cite{flux}  to generate paired-end short reads simulating real RNA-Seq experiment \emph{in silico} based on a ground truth transcript expression profile, using hg19 reference human genome and RefSeq annotations downloaded from UCSC Genome Browser. To generate the ground-truth expression profiles, the gene expressions were sampled from a poisson distribution and the proportions of the isoforms in each gene were derived based on a neighbor average expression in the small transcript network and an initial mixed power law expression profile with gaussian noise. A sequential updating was used to compute the proportion of each isoform by adding the neighbors' average expressions to the initial expression. The update procedure can be found in the S2 Text in {\bf Supplementary.} At last, flux-simulator was applied to simulate the short reads based on the ground truth transcript expression file. 15 million 76-bp paired reads were generated by Flux Simulator and mapped to the reference genome by TopHat \cite{Tophat2} with up to two mismatches allowed. To account for the large dynamic range of abundances, the expressions were normalized by $\log2$(expression+1).

\begin{figure}[H]
\vspace{-1mm}
\centering
\begin{tabular}{c}
{\scalebox{0.78}{\includegraphics*{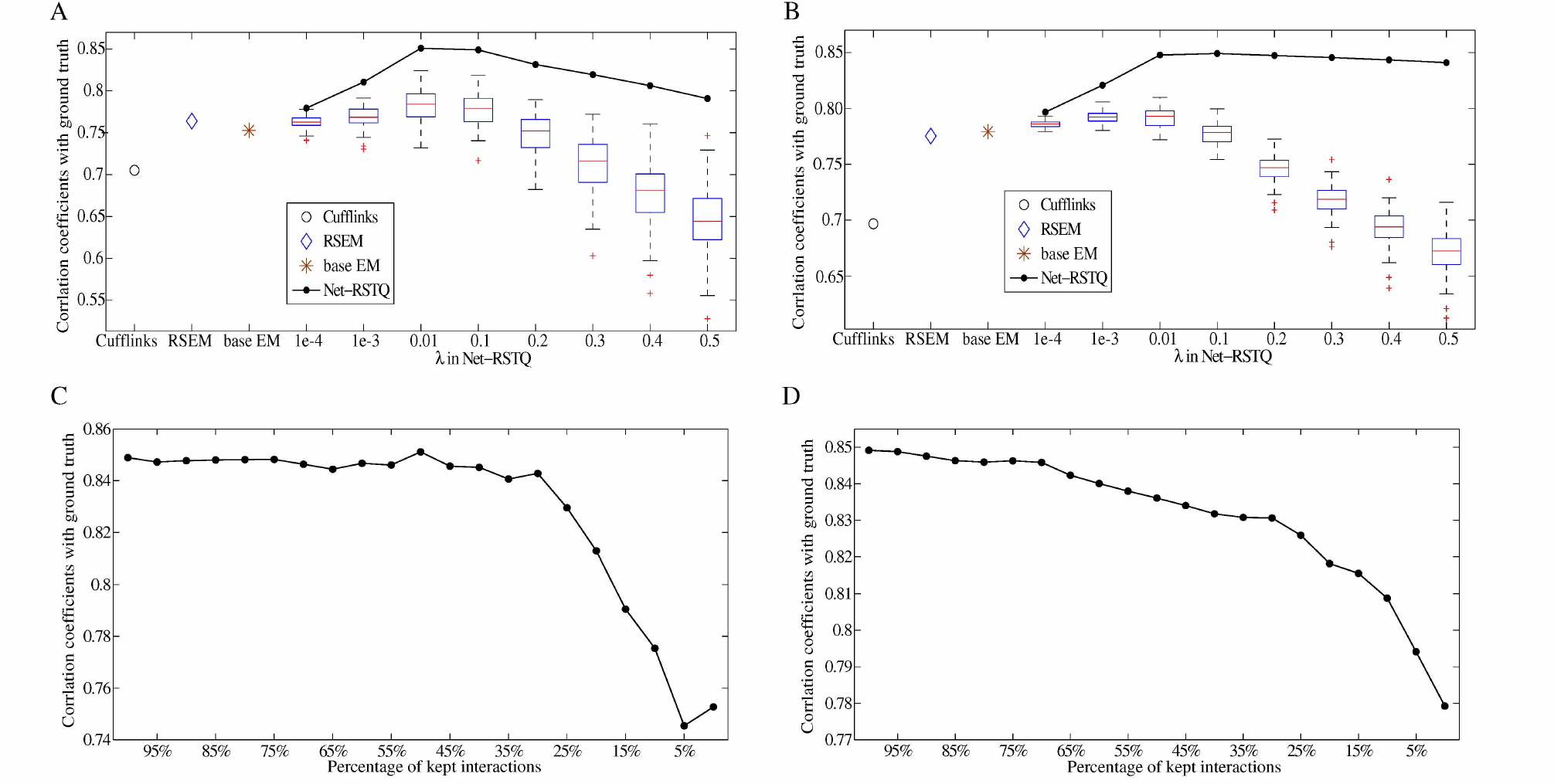}}}
\end{tabular}
\vspace{-1mm}
\caption{{\bf Correlation between estimated transcript expressions and ground truth in simulation.} In (A) and (B) x-axis are labeled by the compared methods and different $\lambda$ parameters of Net-RSTQ. The bar plots show the results of running Net-RSTQ with 100 randomized networks. In (C) and (D), x-axis are are the percentage of edges that are removed from the networks. The plots show the results of running Net-RSTQ with the incomplete networks. (A) and (C)  report the results of 109 transcripts of the isoforms in the same gene with different domain-domain interactions. (B) and (D) report the results of 712 isoforms in genes with multiple isoforms.}\label{fig:simulation}
\vspace{-1mm}
\end{figure}

The correlation coefficients between the transcript abundances estimated by Net-RSTQ under various $\lambda$, base EM (equation \eqref{eqn:likelihood1}), Cufflinks and RSEM, and the ground truth transcript abundances are reported in Fig \ref{fig:simulation}. Furthermore,  Net-RSTQ was also tested with 100 randomized networks with permuted indexes of transcripts in the transcript network.
To assess the impact of the network prior, two cases are shown. Fig \ref{fig:simulation}(A) reports the correlation between the transcripts in which isoforms coded by the same gene are connected with different neighbors (109 out of 898 transcripts in 29 genes). Fig \ref{fig:simulation}(B) reports the results from all the genes with more than one isoform (712 out of 898 transcripts in 211 genes). In both comparisons,  the transcript expressions estimated by Net-RSTQ achieve higher correlation with the ground truth compared with base EM, Cufflinks and RSEM. Slightly higher improvement was observed in the first case than in the second case since the network prior plays more significant role in differentiating the isoform expressions by their different neighbors. When randomized networks are used, Net-RSTQ leads to similar or worse results due to the wrong prior information. Note that since the datasets were generated to partially conform to the network prior, the isoform expressions are relatively ``smooth'' among the neighboring isoforms. Net-RSTQ tends to generate smoother expressions than base EM, Cufflinks and RSEM. When applying Net-RSTQ with small $\lambda$s and randomized network priors, slight improvement was also observed due to the smoothness assumption on the data.

To evaluate the effect of missing edges in the transcript network due to the undetected protein domain-domain interactions, we randomly removed certain percentages of the edges in the transcript network and then run Net-RSTQ with $\lambda=0.1$ on the incomplete networks. The results are shown in Fig \ref {fig:simulation} (C) and (D) for the 109 transcripts with different neighbors and the 712 transcripts in the gene with more than one transcript, respectively. It is intriguing to observe that only when a large percentage of the edges are removed, the performance of Net-RSTQ is affected. Intuitively, the observation can be explained by the fact that the Dirichlet prior parameter is proportional to the average of the neighbors' expressions. As long as some of the neighbors are still connected to the target transcript in the network, the prior information is still useful. The result suggests that Net-RSTQ is relatively robust to utilize transcript networks potentially constructed with a large percentage of undetected protein domain-domain interactions.

\subsection*{Three qRT-PCR experiments confirmed overall improved transcript quantification}
\begin{figure}[!]
\vspace{-1mm}
\centering
\begin{tabular}{c}
{\scalebox{0.9}{\includegraphics*{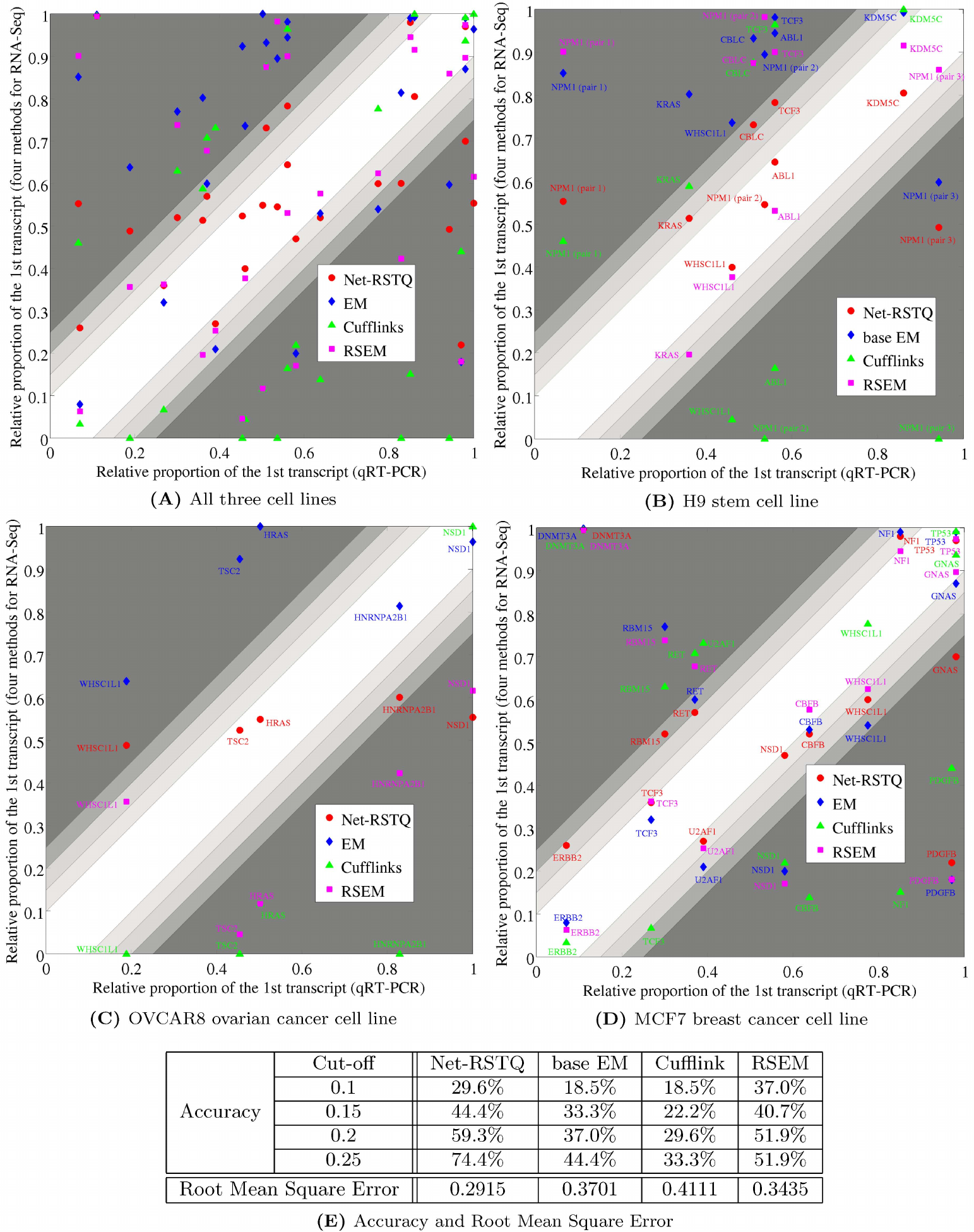}}}
\end{tabular}
\vspace{-1mm}
  \caption{{\bf Validation by comparison with qRT-PCR results.} (A) The scatter plots compare the reported relative proportion of each pair of the isoforms of each gene between the computational methods (Net-RSTQ, base EM, Cufflinks, and RSEM) and qRT-PCR experiments. The proportions of the two compared isoforms in a pair are normalized to adding to 1. The x-axis and y-axis are the relative proportion of one of the two isoform (the other is 1 minus the proportion) reported by qRT-PCR and the computational methods, respectively. The scatter points aligning closer to the diagonal line indicate better estimations by a computational method matching to the qRT-PCR results. The unshaded gradient around the diagonal line shows the regions with scatter differences less than 0.1, 0.15, 0.2 and 0.25, within which the estimations are more similar to the qRT-PCR results. (B)-(D) The scatter plots on each individual dataset. (E) The table shows the percentage of predictions by each method within the unshaded regions and the overall Root Mean Square Error of the predictions by each method compared to the qRT-PCR results.
}\label{fig:Scatter}
\end{figure}
The isoform proportions estimated by Net-RSTQ, base EM, RSEM, and Cufflinks were compared to the qRT-PCR results on the three cell lines.  Parameter $\lambda=0.1$ was fixed in all the Net-RSTQ experiments. Among the genes that Net-RSTQ, base EM, RSEM, and Cufflinks report most different quantification results, qRT-PCR experiments were performed to test the genes with relatively higher coverage of RNA-Seq data, coding two to three isoforms, and the feasibility of designing isoform-specific primers in the qRT-PCR products (see S1, S2 and S3 Tables). Twenty-five genes in total were tested in the three cell lines: seven in H9 stem cell line, five in OVCAR8 ovarian cancer cell line, and thirteen in MCF7 breast cancer cell line. The scatter plots of the relative abundance of the first transcript in each gene estimated by Net-RSTQ, base EM, Cufflinks and RSEM were compared to the qRT-PCR results in Fig \ref{fig:Scatter}(A) and (E). In the scatter plot, the estimated relative abundance by Net-RSTQ were closer to qRT-PCR results measured by the accuracy of various thresholds and Root Mean Square Errors.
Net-RSTQ achieved the lowest Root Mean Square Error of 0.291, which is more than 0.05 less than 0.3435, the second best achieved by RSEM. In the 20\% confidence region, Net-RSTQ puts 59.3\% of the pairs in the region compared with 37\%, 29.6\%, and 51.9\% by base EM, Cufflink, and RSEM, respectively. RSEM performed well by putting 37.0\% of the pairs within 10\% confidence regions but performed poorly in about half of the pairs with more than 25\% error.

The relative abundance of the seven genes in H9 stem cell line is shown in Fig \ref{fig:Scatter}(B), S2(A) Fig and S5 Table. In all seven genes tested, the relative abundance estimated by Net-RSTQ is closer to the qRT-PCR results compare to that by base EM and Cufflinks. RSEM performed similarly well on four genes and worse on the other three genes, CBLC, TCF3 and NPM1. The same comparison on the five selected genes in OVCAR8 ovarian cancer cell line is shown in Fig \ref{fig:Scatter}(C), S2(B) Fig and S6 Table. Cufflinks reports very low expressions in the first transcript in four genes, three of which do not agree with the highly expressed transcript in the qRT-PCR results. While base EM performed better for two genes (NSD1 and HNRNPA2B1), Net-RSTQ performed better on the other three genes (HRAS, TSC2, and WHSC1L1). Net-RSTQ correctly predicted the overall enrichment of isoforms of HNRNPA2B1 and NSD1 (NM$\_$031243 $>$ NM$\_$002137 in HNRNPA2B1 and NM$\_$022455 $>$ NM$\_$172349 in NSD1). It is possible that the expressions of NM$\_$002137 transcript in gene HNRNPA2B1 and NM$\_$172349 in gene NSD1 were slightly over-smoothed by network information in Net-RSTQ with the fixed $\lambda$ parameter. RSEM performed slightly better on WHSC1L1 and NSD1 but much worse in the other three genes. The same comparison on the thirteen genes in MCF7 breast cancer cell line is shown in Fig \ref{fig:Scatter}(D), S2(C) Fig and S7 Table. Cufflinks performed poorly on 8 genes with more than 25\% error while RSEM, base EM and Net-RSTQ performed poorly on 5, 4 and 3 genes, respectively. Overall, Net-RSTQ performed better than base EM and Cufflinks and slightly better than RSEM.
In summary, Net-RSTQ improved the overall isoform quantification significantly in the H9 stem cell data and predicted more consistent cases in OVCAR8 and MCF7 cancer cell lines data. Note that there could be more uncertainties in primer designs due to somatic DNA variations and cell differentiation and proliferation in cancer cell lines, potentially a larger variation in the qRT-PCR experiments on the cancer cell lines is expected than H9 stem cell line.

\subsection*{Net-RSTQ improved overall cancer outcome predictions}

To provide an additional evaluation of the quality of transcript quantification, we designed six cancer outcome prediction tasks by the assumption that better transcript quantification always leads to better isoform markers for cancer outcome prediction. Net-RSTQ was compared with base EM, RSEM \cite{RSEM}, and Cufflinks \cite{cufflinks} by classification with the quantification of isoform transcripts in two cancer gene lists (397 and 2551 genes) on four cancer datasets. Each dataset is divided into four folds with two folds for training, one fold for validation (parameter tuning), and one fold for test in a four-fold cross-validation. Support Vector Machine (SVM) with RBF kernel \cite{SVMRBF} were chosen as the classifier. We repeated the four-fold cross-validation 100 times by each method in each dataset.

The average area under the curve (AUC) of receiver operating characteristic of the 100 repeats are reported in Table \ref{tab:classification} when the small gene list was used and Table \ref{tab:classification2} when the large gene list was used. The transcript expressions estimated by Net-RSTQ consistently achieved better average classification results than those by the base EM. To evaluate the statistical significance of the differences between the AUCs generated by Net-RSTQ and the base EM in the 100 repeats, we also report the $p$-values by a binomial test on the number of wins/loses in all the experiments between Net-RSTQ and the base EM in Table \ref{tab:classification} and Table \ref{tab:classification2}.
When the small gene list was tested, three cases were significant with low $p$-values less than $0.001$ and two cases were significant with $p$-values just below $0.02$ while in the BRCA (survival) data, the $p$-value is only moderately significant even though the average by Net-RSTQ is higher. Overall, Net-RSTQ outperformed the base EM significantly. When the larger gene list was tested, the improvements are not as significant. The improvement was only significant in one dataset, BRCA (survival), and slightly significant in two datasets, OV (relapse) and LUSC (survival). In the other three datasets, the improvements are not significant. Net-RSTQ also outperformed Cufflinks and RSEM (transcript or gene) in five cases except the experiment on BRCA (relapse) dataset in Table \ref{tab:classification}. In Table \ref{tab:classification2}, the improvements are less obvious. Moreover, the isoform expression features are not more informative than gene expression features. Overall, the classification performance with the small gene list in Table \ref{tab:classification} is generally better than or similar to the large gene list in Table \ref{tab:classification2} possibly suggesting less relevance to survival and relapse in the large gene list.

\begin{table}[H]
\tiny
\centering
\scalebox{.9}{
\begin{tabular}{|c|c|c|c|c|c|c|}
\hline
{\bf Dataset}&{\bf OV(Survival)}&{\bf OV(Relapse)}&{\bf BRCA(Survival)}&{\bf BRCA(Relapse)}&{\bf LUAD(Survival)}&{\bf LUSC(Survival)}\\
\hline
{\bf Net-RSTQ(Isoform)}&{\bf 0.597}&{\bf 0.607}&{\bf 0.683}&{0.590}&{\bf 0.635}&{\bf 0.567}\\
{\bf base EM(Isoform)}&{0.570}&{0.589}&{0.673}&{0.542}&{0.579}&{0.550}\\
{\bf RSEM(Isoform)}&{0.587}&{0.550}&{0.651}&{\bf 0.616}&{0.613}&{0.536}\\
{\bf Cufflinks(Isoform)}&{0.563}&{0.577}&{0.676}&{0.593}&{0.555}&{0.556}\\
\hline
{\bf RSEM(Gene)}&{0.591}&{0.580}&{0.651}&{0.558}&{0.615}&{0.559}\\
\hline
{\bf p-value(Net-RSTQ vs base EM)}&{0.0011}&{0.0198}&{0.1356}&{2.248e-5}&{1.948e-8}&{0.0167}\\
\hline
\end{tabular}}
\caption{{\bf Classification performance of estimated transcript expressions and gene expression on the small cancer gene list.} The mean AUC scores of classifying patients by estimated transcript (gene) expression in four-fold cross-validation for each dataset are reported. The best AUCs across the five models using isoforms as features are bold.} \label{tab:classification}
\end{table}
\begin{table}[H]
\tiny
\centering
\scalebox{.9}{
\begin{tabular}{|c|c|c|c|c|c|c|}
\hline
{\bf Dataset}&{\bf OV(Survival)}&{\bf OV(Relapse)}&{\bf BRCA(Survival)}&{\bf BRCA(Relapse)}&{\bf LUAD(Survival)}&{\bf LUSC(Survival)}\\
\hline
{\bf Net-RSTQ(Isoform)}&{0.599}&{\bf 0.585}&{0.679}&{0.592}&{0.604}&{\bf 0.566}\\
{\bf base EM(Isoform)}&{0.590}&{0.572}&{0.651}&{0.571}&{0.597}&{0.556}\\
{\bf RSEM(Isoform)}&{0.584}&{0.569}&{0.663}&{0.594}&{0.587}&{0.543}\\
{\bf Cufflinks(Isoform)}&{0.562}&{0.582}&{\bf 0.683}&{0.580}&{0.583}&{0.559}\\
\hline
{\bf RSEM(Gene)}&{\bf 0.604}&{0.577}&{0.675}&{\bf 0.598}&{\bf 0.627}&{0.554}\\
\hline
{\bf p-value(Net-RSTQ vs base EM)}&{0.3798}&{0.0967}&{0.0018}&{0.3822}&{0.6178}&{0.1356}\\
\hline
\end{tabular}}
\caption{{\bf Classification performance of estimated transcript expressions and gene expression on the large cancer gene list.} The mean AUC scores of classifying patients by estimated transcript (gene) expression in four-fold cross-validation for each dataset are reported. The best AUCs across the five models are bold.} \label{tab:classification2}
\end{table}

The parameter $\lambda$ was tuned by the AUC on the validation set and the optimal $\lambda$ was used to train the Net-RSTQ model to be tested on the test set. The process is repeated for each fold in 100 repeats. To show the effect of varying the $\lambda$ on the classification performance in Net-RSTQ, we plotted the average AUC on the validation set across the 100 repeats on the BRCA (survival) dataset with small gene list in S3(A) Fig. The optimal $\lambda$ was 0.1 in this experiment. The local gradient around the optimal $\lambda$ suggesting that the transcript network is playing an important role in inferring better transcript quantification from the RNA-Seq data. In S3(B) Fig, the convergence of Net-RSTQ is also illustrated by each update through all the genes in each iteration. After less than 10 overall iterations across 397 genes, Net-RSTQ converged well to a local optimum. Similar convergence patterns were observed in all other TCGA samples.

\begin{figure}[H]
\vspace{-1mm}
\centering
\begin{tabular}{c}
{\scalebox{0.8}{\includegraphics*{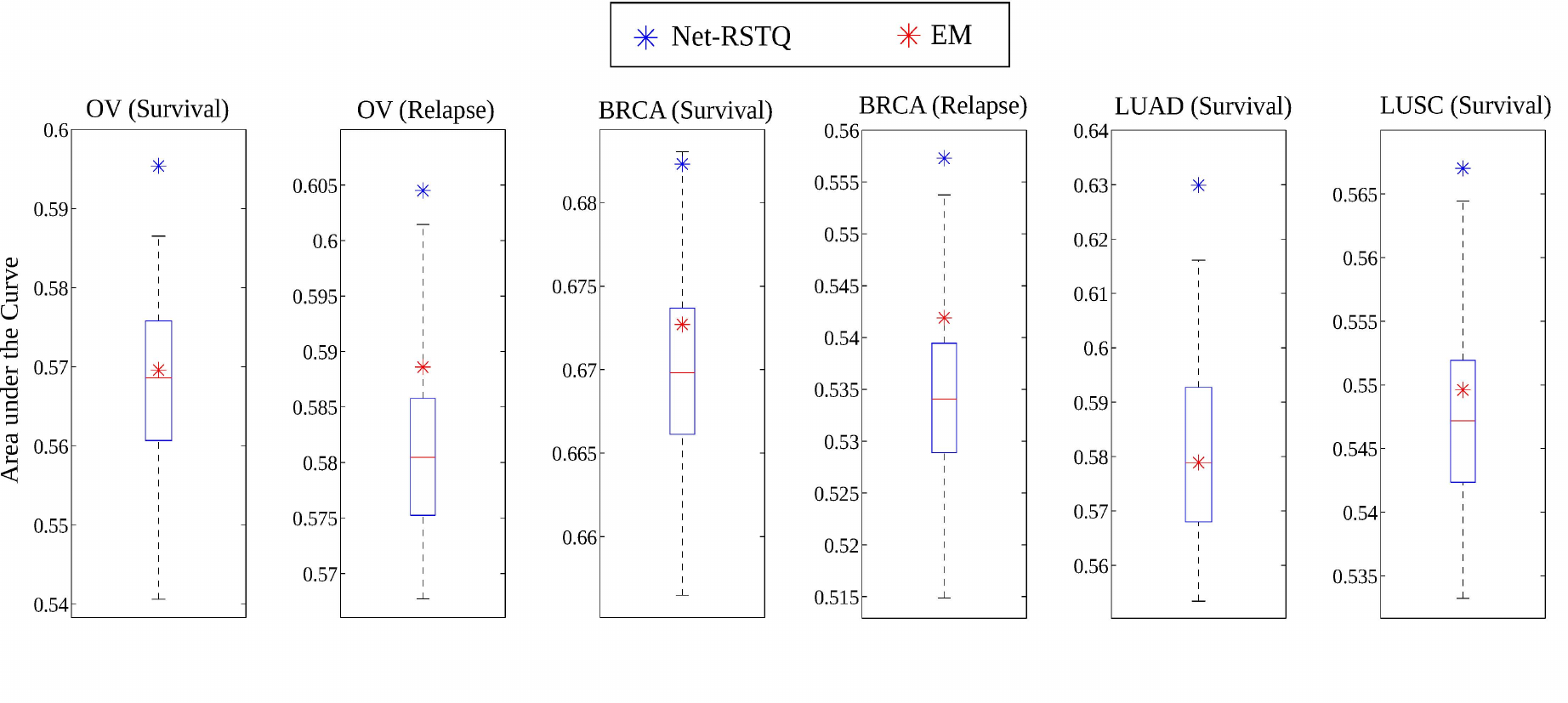}}}
\end{tabular}
\vspace{-1mm}
\caption{{\bf Statistical analysis with randomized networks.} Comparison of the classification results by the randomized networks and the true network. The $\lambda$ parameter was fixed to be 0.1 in all the experiments. The blue star and the red star represent the results with the real network and without network (base EM), respectively. The boxplot shows the results with the randomized networks.}\label{fig:randomization}
\vspace{-1mm}
\end{figure}

To understand the role of the transcript network in the transcript expression estimation, we used 100 randomized networks to learn the transcript proportion in each experiment with $\lambda$ fixed to be 0.1. In each randomization, the edges were shuffled among all the transcripts in the small gene list. For transcript expressions learned by each randomized network, we conducted the same four-fold cross validation to compute the average AUCs among 100 repeats. The boxplot of the AUCs learned with the 100 randomized networks is shown in Fig \ref{fig:randomization}. Compared with the classification results from the true transcript network, the result with randomized networks is always worse. Another important observation is that, the median value of the AUCs across the 100 randomized networks is lower or close to the result by the base EM, which suggests that the randomized networks play no role in improving classification and even lead to worse result. Overall, the results provide a clear evidence that the transcript network is informative for the transcript expression estimation, and supplies more discriminative features for cancer outcome prediction.
\subsection*{Running time}
\begin{figure}[H]
\vspace{-1mm}
\centering
\begin{tabular}{c}
{\scalebox{0.8}{\includegraphics*{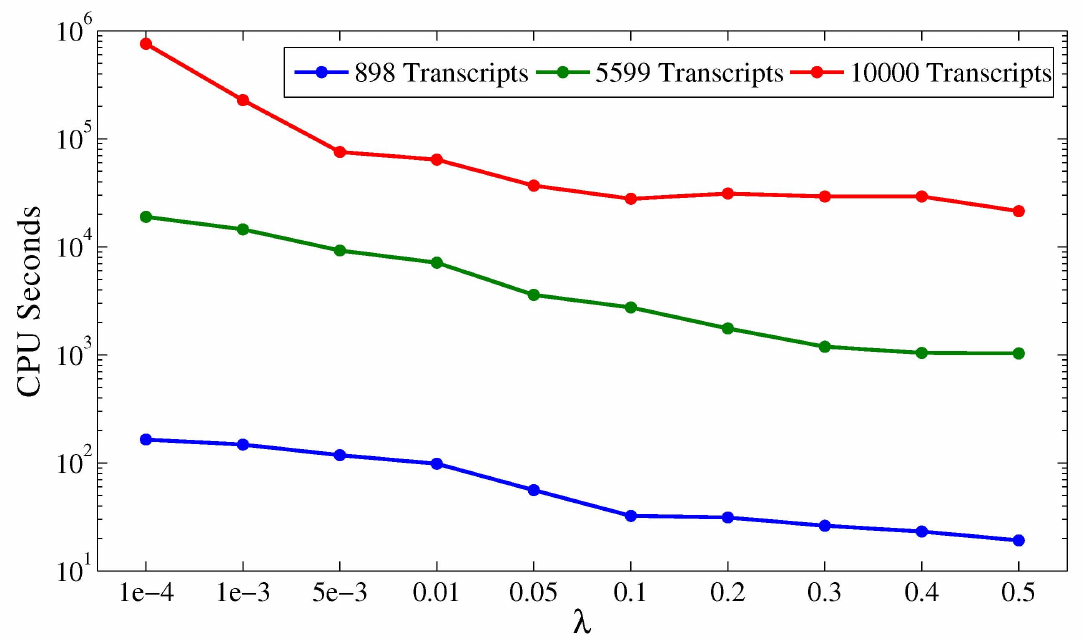}}}
\end{tabular}
\vspace{-1mm}
\caption{\bf{Running time.} The plots show the CPU time (Intel Xeon E5-1620 with 3.70GHZ) for running the Net-RSTQ algorithm one three networks, the small transcript network, the large transcript network, and an artificial huge network of 10000 transcripts.}\label{fig:cpu}
\vspace{-1mm}
\end{figure}

To measure the scalability of Net-RSTQ, we tested the Net-RSTQ algorithm on the data of the MCF7 breast cancer cell line with three different networks, the small network (898 transcripts), the large network (5599 transcripts) and an artificial huge network (10000 transcripts).
Fig \ref{fig:cpu} plots the CPU seconds of running Net-RSTQ on the three networks under different $\lambda$s.  On the small network, the running time is at most about 100 seconds while on the large network and the huge network, the running time is in the scale of 1-$e^3$$\sim$1-$e^4$ and 1-$e^5$$\sim$1-$e^6$, respectively. When $\lambda=0.1$, the CPU time for the small network is 32.4 seconds; for the large network is 2755 seconds; and for the artificial large network is 27806 seconds. The results suggest that Net-RSTQ might scale up to about 10000 transcripts, and thus the performance is sufficient for studies focusing on any pathway with up to several thousand genes in the pathway.
\section*{Discussion}
In the paper, we explored the possibility of improving short-read alignment based transcript quantification with relevant prior knowledge, protein domain-domain interactions. The observation of the correlation between isoform co-expressions and protein domain-domain interactions suggests that the approach is a well-grounded exploration. Different from previously methods \cite{RNASeqReview}, Net-RSTQ is a network-based approach that directly incorporates protein domain-domain interaction information for transcript proportion estimation. The experiments suggested a great potential of exploring protein domain-domain interactions to overcome the limitations of short-read alignments and improve transcript quantification for better sample classification.

The Dirichlet prior from the neighboring isoforms play two different roles: differentiating isoform expressions to reflect different functional roles or smoothing isoform expressions to reflect similar functional roles, depending on whether the isoforms of a gene share the same or different interacting partners. This principle in modeling is based on the hypothesis that isoforms playing different functional roles (e.g. containing different protein domains) are more likely to behavior differently than isoforms with the same or similar functional roles (e.g. containing the same protein domains). When the isoforms of a gene interact with different partners, their expressions correlates with their partners' expressions. And, when the isoforms of a gene interact with the same partners, there is no benefit on differentiating their proportions to drive the functionality. A limitation is that when the functional difference among the isoforms are not captured by domain content, the smoothing role might under-estimate the difference in their proportions. Thus, our future goal is to bring in other type of functional information to distinguish their functional roles in cancer such as preferential adoption of post-transcriptional regulations.

Currently, Net-RSTQ does not directly model multi-hits reads in multiple loci. In the TCGA experiments, around 5-10\% of the aligned reads in four datasets have multiple alignments reported by TopHat and only one of the best alignments is considered. To check the effect of the multiple-alignment reads in transcript quantification, we allow up to 20 best alignments by TopHat and normalized the read assignment $q_{ijk}$ by the number of loci that the reads aligned to. The correlation coefficients between the estimated gene expressions before and after the normalization are above 0.98 in all the datasets. A potential rigorous solution is to add iteratively reassignment of the reads to the potential origins based on updated abundance of the involved isoforms. The modification will significantly decrease the computational efficiency and make it impractical on large RNA-Seq datasets.

There is also another alternative of integrating the network information directly as a regularization term on the joint likelihood function of all the genes. We also explored this model in the S1 Text in \textbf{Supplementary}. In the preliminary experiments, we observed very similar outputs between the alternative model and the Net-RSTQ model shown in S8 Table. However, since the alternative model directly works with one large optimization problem across all the genes, the convergence is much slower as shown in S4 Fig and the optimization package used in the experiments ran into numerical issues. Thus, we believe the Net-RSTQ model is more scalable and robust in comparison.

Currently, Net-RSTQ can scale on transcript network with up to around 5000 transcripts, which is sufficient for more focused analysis of several thousand genes. The running time of Net-RSTQ on such large transcript network is below 2 hours on each TCGA sample, compared with 5-8 hours needed for aligning the short reads. To further scale up Net-RSTQ, we will investigate other faster strategies of utilizing short read information, such as Sailfish \cite{Sailfish} which directly estimates isoform expressions by counting k-mer occurrences in reads rather than reads from the alignments. This will be our future direction.


\section*{Acknowledgments}
The results are based upon data generated by The Cancer Genome Atlas established by the NCI and NHGRI. Information about TCGA and the investigators and institutions who constitute the TCGA research network can be found at \url{http://cancergenome.nih.gov}. The dbGaP accession number to the specific version of the TCGA dataset is phs000178.v8.p7.


%
%
%

\bibliography{bioinfo2}
\newpage
\section*{Supporting Information}
{\bf S1 Text. Alternative model by network-based regularization.}
We introduce a network-based regularization to the base model as an alternative model and evaluate the probabilities of a read being generated by the transcripts in all the genes simultaneously as follows,
\begin{eqnarray}
\mathcal{L}_{pen}({\bm P};{\bm r})= log(\mathcal{L}({\bm P};{\bm r}))-\lambda\|\bm{AP}-\bm{WP}\|_{2}^{2}.\label{eqn:PenLike}
\end{eqnarray}
The term $\lambda\|\bm{AP}-\bm{WP}\|_{2}^{2}$ in equation \eqref{eqn:PenLike} is a network constraint to encode prior knowledge from the transcript network. Given a transcript interaction network, we assume that the connected transcripts are more likely to co-express by introducing the following cost term over the expression $\bm{\pi}$,
\begin{eqnarray}
\Psi(\bm{P,\pi}) &=& \sum^{|\bm{T}|}_{i=1}\left(\pi_{i}-\sum_{j\in\bm{nb}(i)}\frac{\pi_{j}}{|\bm{nb}(i)|}\right)^2\nonumber\\
&=&\sum^{|\bm{T}|}_{i=1}\left(\frac{p_i|\bm{r}_{g(i)}|}{l_i}-\sum_{j\in\bm{nb}(i)}\frac{p_j|\bm{r}_{g(j)}|}{|\bm{nb}(i)|l_j}\right)^{2}\nonumber\\
&=&\sum^{|\bm{T}|}_{i=1}\left(A_{ii}p_i-\sum_{j\in\bm{nb}(i)}W_{ij}p_j\right)^2\nonumber\\
&=&\|\bm{AP}-\bm{WP}\|_{2}^{2},
\end{eqnarray}
where $\bm{nb}(i)$ are the neighbors of transcript $i$. $g(i)$ and $g(j)$ are the genes containing transcripts $i$ and $j$, respectively. $|\bm{r}_{g(i)}|$ denotes the number of reads aligned to gene $g(i)$. $l_i$ and $l_j$ are the length of transcripts $i$ and $j$. $\bm{A}$ is a diagonal matrix, where $A_{ii}=|\bm{r}_{g(i)}|/l_{i}$. $\bm{W}$ contains the weights of transcript pairs in the transcript network, where $W_{ij} = |\bm{r}_{g(j)}|/(|\bm{nb}(i)|l_j)$. Minimizing $\Psi(\bm{P,\pi})$ ensures that each transcript will receive an expression close to the average expression of its neighbors in the transcript network. To solve equation \eqref{eqn:PenLike} we used CVX, a package for specifying and solving convex programs \cite{cvx,gb08}. The framework estimates the expressions of transcripts in all the genes together in one optimization. We applied this framework to the small network with 898 transcripts on MCF7 breast cancer cell line RNA-Seq data. Overall, the results between Net-RSTQ and the alternative framework can be highly similar as shown in S8 Table when parameter ${\lambda}$s are tuned. However, the algorithm converges slowly (S4 Figure) compared to the convergence of Net-RSTQ in Figure 7 in the main manuscript). It is clear that the alternative model does not scale to larger networks.
\vspace{3mm}

\noindent{\bf S2 Text. Steps of generating the simulation data.}
Below are the steps of generating the simulation data,

\begin{enumerate}
\item Generate 397 gene expression $E$ by sampling a poisson distribution.
\item For the genes with multiple transcripts, generate isoform expressions from a mixed power law by the flux-simulator to calculate the initial proportion $p_{ik}(0)$ for each transcript $T_{ik}$.
\item Let the initial isoform expression $\pi_{ik}(0) = E_i*p_{ik}(0)$ + gaussian noise
\item Let $\alpha = 1$; Repeat  $p_{ik}= (\alpha* \sum_{j \in nb(T_{ik})} \frac{\pi_{g(j),j}}{|nb(T_{ik})|} + \pi_{ik}(0)) / \sum_q (\alpha* \sum_{j \in nb(T_{iq})} \frac{\pi_{g(j),j}}{{|nb(T_{iq})|}} + \pi_{iq}(0))$.
\item After convergence, $\pi_{ik} = E_i*p_{ik}$.
\item ${\bm \pi}$ is further normalized and used with flux-simulator as the ground truth expressions.
\end{enumerate}
\vspace{3mm}

\noindent{\bf S3 Text. Cufflinks and RSEM command line.}
Cufflinks 2.1.1\cite{cufflinks} was applied to generate isoform expression as one of the baseline to compare with Net-RSTQ using the following command:

  {\bf./cufflinks -p 4 -F 0 -G hg19RefSeq.gtf -o x.bam}.

\noindent RSEM1.2.20\cite{RSEM} was applied to generate isoform expression as another baseline to compare with Net-RSTQ in simulation and qRT-PCR studies using the following command:

  {\bf./rsem-calculate-expression --paired-end --bowtie2 --bowtie2-path bowtie2/ -p 2 x1.fastq x2.fastq hg19RefSeq x}.
 \vspace{3mm}

\begin{figure}[H]
\vspace{-1mm}
\centering
\begin{tabular}{c}
{\scalebox{0.8}{\includegraphics*{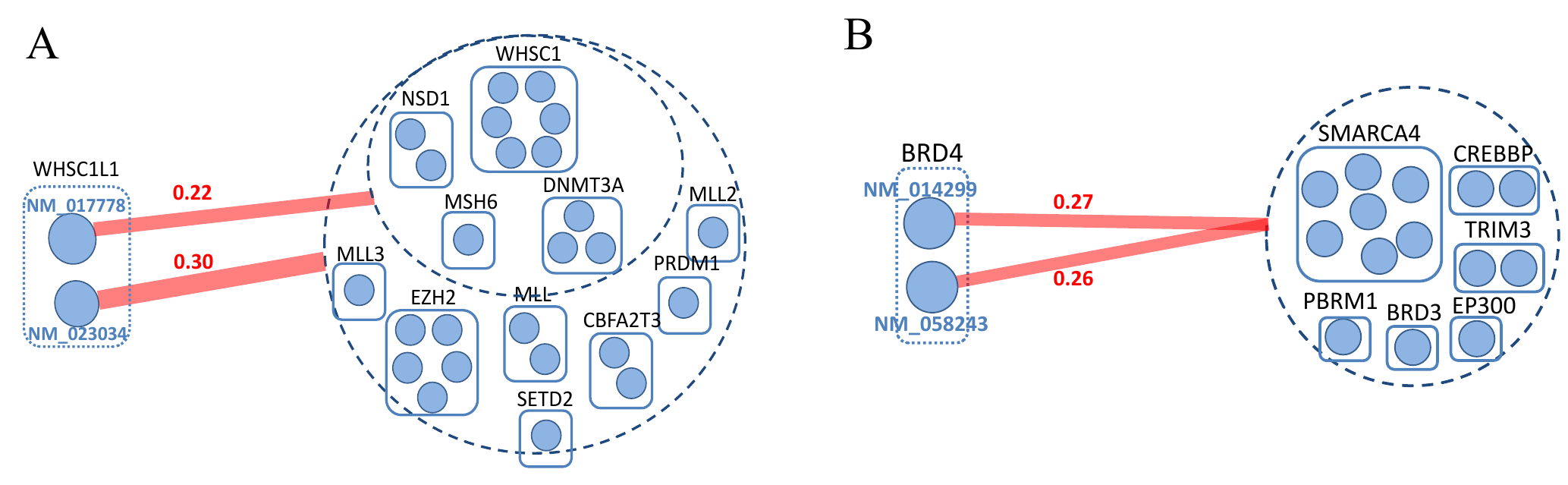}}}
\end{tabular}
\vspace{-1mm}
\end{figure}
\noindent{\bf S1 Figure. Examples of transcript sub-networks with co-expression information.} (A) Transcripts in WHSC1L1 with correlation coefficients calculated on the OV dataset. (B) Transcripts in BRD4 with correlation coefficients calculated on the BRCA dataset.  Both examples are shown with the neighbors in the small transcript network.
\vspace{4mm}

\begin{figure}[H]
\vspace{-1mm}
\centering
\begin{tabular}{c}
{\scalebox{0.65}{\includegraphics*{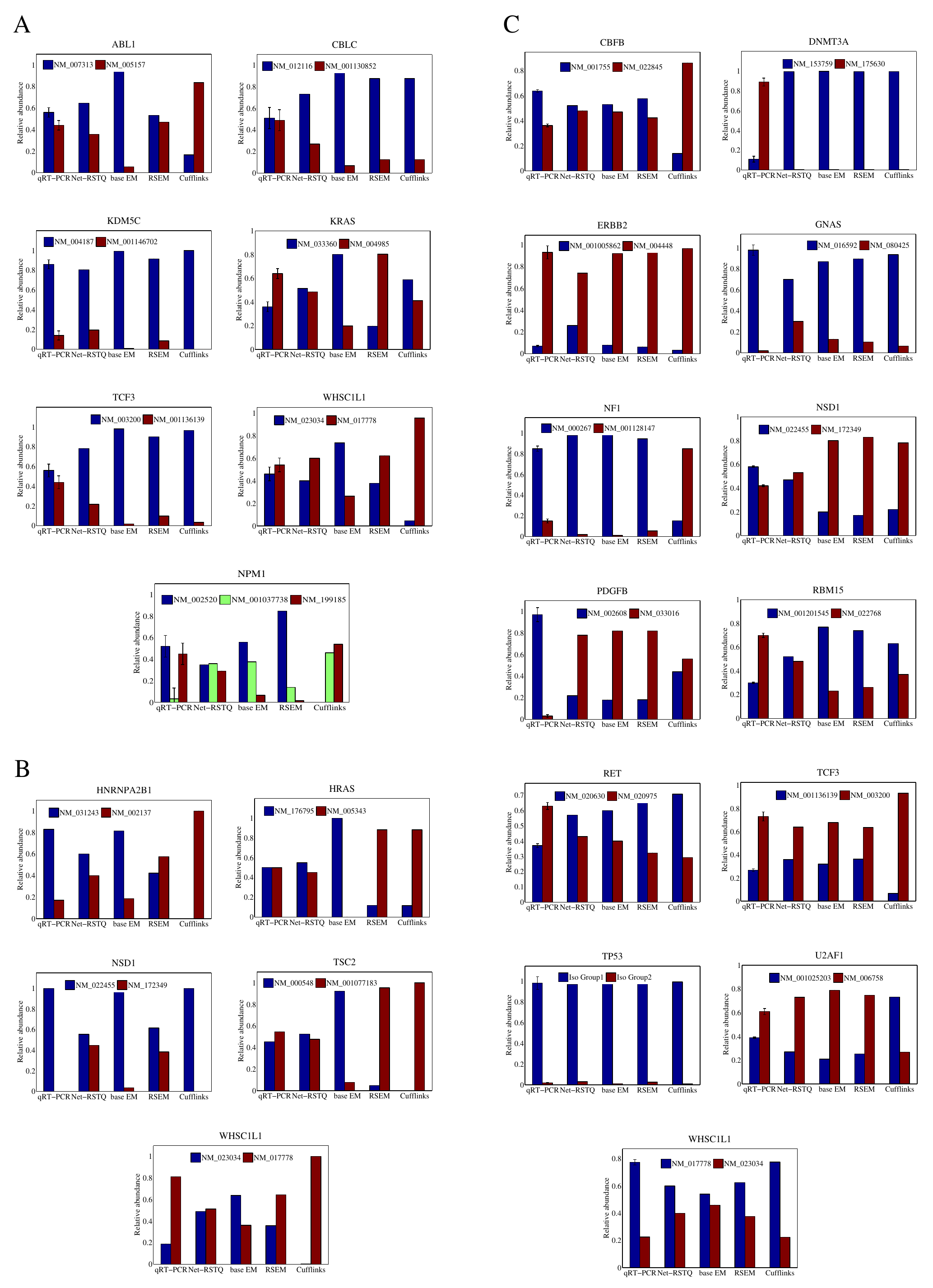}}}
\end{tabular}
\vspace{-1mm}
\end{figure}
\noindent{\bf S2 Figure. Evaluation by qRT-PCR experiments.} The relative abundance of the transcripts in 7 tested genes in H9 stem cell line (A), 5 tested genes in OVCAR8 ovarian cancer cell line (B), and 13 tested genes in MCF7 breast cancer cell line (C) estimated by Net-RSTQ, base EM, Cufflinks and RSEM was compared with the qRT-PCR experiments. The total abundance is normalized to 1 over the measured transcripts in each gene.
\vspace{4mm}

\begin{figure}[H]
\vspace{-1mm}
\centering
\begin{tabular}{c}
{\scalebox{0.5}{\includegraphics*{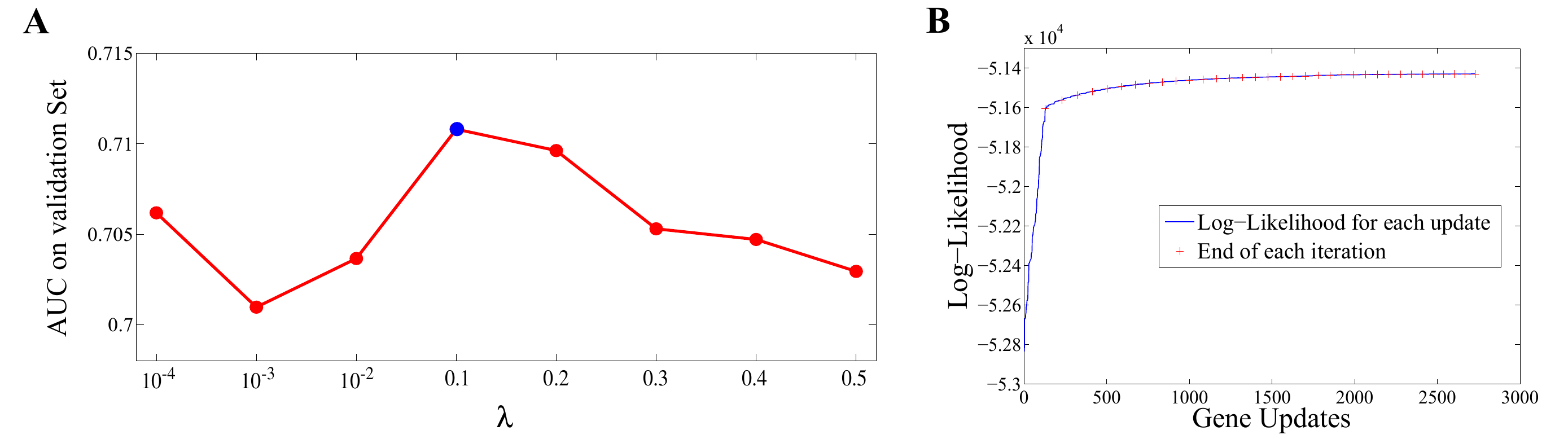}}}
\end{tabular}
\vspace{-1mm}
\end{figure}
\noindent{\bf S3 Figure. Model selection and convergence.} The experiment done on BRCA (survival) dataset. (A) Effect of varying $\lambda$ on the classification performance. The plot shows the average AUC learned from the 100 repeats on validation set for different $\lambda$s with the optimal $\lambda$ in blue. (B) Convergence analysis by the total log-likelihood. The plot shows the change of total log-likelihood in Net-RSTQ with each gene update. Each red cross indicates the end of each round $t$ in line 2 of Algorithm 1.
\vspace{4mm}

\begin{figure}[H]
\vspace{-1mm}
\centering
\begin{tabular}{c}
{\scalebox{0.5}{\includegraphics*{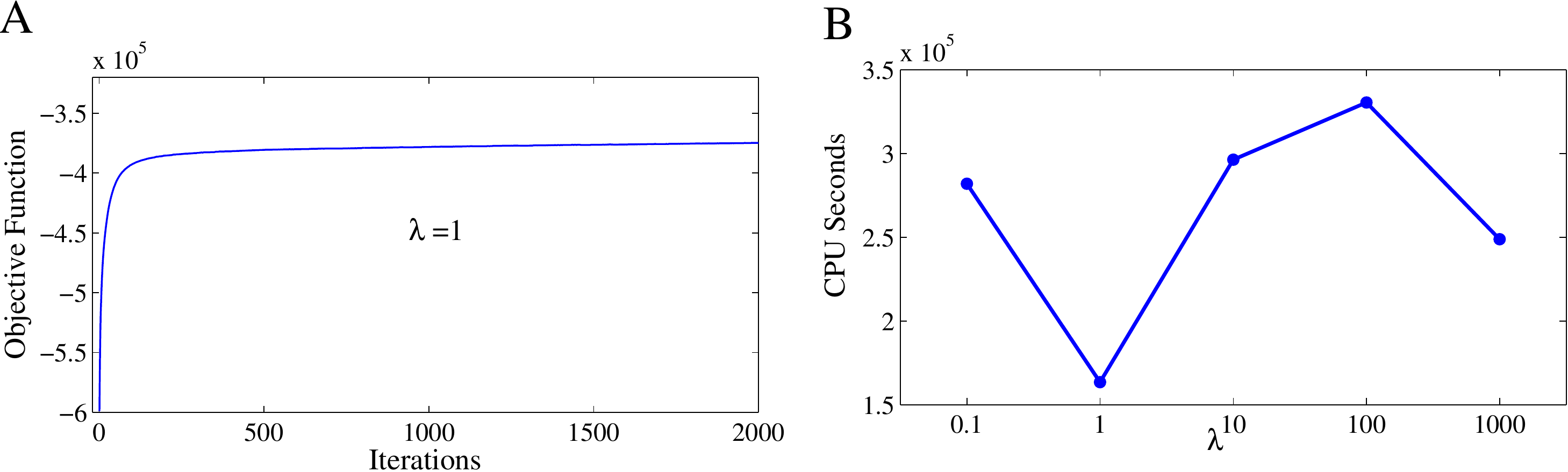}}}
\end{tabular}
\vspace{-1mm}
\end{figure}
\noindent{\bf S4 Figure. (A) Convergence and (B) Running time of the alternative regularized framework with 2000 iterations on MCF7 breast cancer cell line.}
\vspace{4mm}
\begin{table}[H]
\tiny
\centering
\begin{tabular}{|c|c|c|c|}
\hline
{Gene/Transcript(isoform) Names}&{Primer Names*}&{Forward}&{Reverse}\\
\hline
{ABL1}&{Template 1\&2}&{5-GGTTGGTGACTTCCACAGGAAA}&{5-CACCGTCAGGCTGTATTTCTTCC}\\
{NM\_007313(iso2)}&{qPCR 1\&2}&{5-TGAAAAGCTCCGGGTCTTAGG}&{5-TTGACTGGCGTGATGTAGTTG}\\
{NM\_005157(iso1)}&{qPCR 2}&{5-TAGCCAAAGACCATCAGCGTT}&{5-TTCGCGGTTATCAATTTCATGT}\\
\hline
{CBCL}&{Template 1\&2}&{5-ACCCTGTGGAACCAGGCTGC}&{5-CACCTGCCCCAGCTCCAACT}\\
{NM\_012116(iso1)}&{qPCR 1\&2}&{5-ACCACCATTGACCTCACCTGC}&{5-ACTGCCAGGAGCTGCCAGTT}\\
{NM\_001130852(iso2)}&{qPCR 1}&{5-CATCCTGCAGACCATCCCTG}&{5-GGCCGAGCTCAGTCAGGTCT}\\
\hline
{KDM5C}&{Template 1\&2}&{5-GACCTGCTCGAGGTGACCCT}&{5-AAGCTTTCTTCAGATCACAGGGAG}\\
{NM\_004187(iso1)}&{qPCR 1\&2}&{5-GCCTCTAACCAGCATTCCCA}&{5-TCTCTGGAATGGTGATGGCC}\\
{NM\_001146702(iso2)}&{qPCR 1}&{5-AGAGGCTGAGGAGGTCCAGG}&{5-CCAAGCCATTCTGGTTCTCC}\\
\hline
{TCF3}&{Template 1\&2}&{5-TGAATCCCAAAGCAGCCTG}&{5-TCTTGTAACTAATGTTTTTATTTTCCTTA}\\
{NM\_003200(iso1)}&{qPCR 1\&2}&{5-TGAATCCCAAAGCAGCCTGT}&{5-GGTTGTGGGCTTCGCTCAG}\\
{NM\_001136139(iso2)}&{qPCR 1}&{5-GTATGCCTCCGTGGGACGA}&{5-GGAGCTCCTGGACCCAGTGT}\\
\hline
\multirow{4}{*}{\begin{tabular}[x]{@{}c@{}}WHSC1L1\\NM\_023034(iso1)\\NM\_017778(iso2)\end{tabular}}&{Template 1}&{5-CAGTTCCTCAGGCTACAGTGAAGA}&{5-CATACAACAAACAGACATCTAGATCAAC}\\
&{Template 2}&{5-CAGTTCCTCAGGCTACAGTGAAGA}&{5-GTAATGTAGTTTCTTGCCAGCTTTACA}\\
&{qPCR 1}&{5-GTCGGCGGCTTGATAAACAGT}&{5-GTACCCATCCAGCTCAAACCG}\\
&{qPCR 2}&{5- CCCTTCAGCTACTGCAGATGC}&{5-CCAGGCACTCCAGGTGAAAGT}\\
\hline
\multirow{4}{*}{\begin{tabular}[x]{@{}c@{}}KRAS\\NM\_033360(iso2)\\NM\_004985(iso1)\end{tabular}}&{Template 1}&{5-TTCCTGCTCCATGCAGACTGT}&{5-TAAGAAGTAATCAACTGCATGCACCA}\\
&{Template 2}&{5-TACATTGGTGAGGGAGATCCGA}&{5-TAAGAAGTAATCAACTGCATGCACCA}\\
&{qPCR 1}&{5-TTCCTGCTCCATGCAGACTGT}&{5-GCACCAAAAACCCCAAGACAG}\\
&{qPCR 2}&{5- TACATTGGTGAGGGAGATCCGA}&{5-TAGAAGGCATCATCAACACCCA}\\
\hline
{NMP1}&{Template 1\&2\&3}&{5-TCCTTTCCCTGGTGTGATTCC}&{5-CATTGTCAGGTGAGGCAAATGC}\\
{NM\_002520(iso1)}&{qPCR 1\&2\&3}&{5-TCCTTTCCCTGGTGTGATTCC}&{5-TCGGCCTTTAGTTCACAACCG}\\
{NM\_0010337738(iso3)}&{qPCR 1\&3}&{5-AGCTGAAGAAAAAGCGCCAGT}&{5-CTTTTGTGCATTTTTGGCTGG}\\
{NM\_199185(iso2)}&{qPCR 3}&{5-AAGCCCAAAGATGGGGAGAA}&{5-AAGGGCAAGGTTCACTGAATCA}\\
\hline											
\end{tabular}
\end{table}
\noindent{\bf S1 Table. Primer sets of the transcripts in seven genes of H9 stem cell line.} * The numbers refer to the isoforms in the first column.
\vspace{4mm}
\begin{table}[H]
\tiny
\centering
\begin{tabular}{|c|c|c|c|}
\hline
{Gene Name}&{Transcript Name}&{Primer Sequence - Forward}&{Primer Sequence - Reverse}\\
\hline
\multirow{2}{*}{HNRNPA2B1}&{NM\_031243}&{TCC GCG ATG GAG GAA AAC TTT AG}&{GCC ACC AAT AAA GAG CTT ACG G}\\
\cline{2-4}
&{NM\_002137}&{AGC GGC AGT TCT CAC TAC AG}&{TCC TTT TCT CTC CTC CAT CG}\\
\hline
\multirow{2}{*}{HRAS$^{*}$}&{NM\_176795}&{CCG CTC TGG CTC TAG CTC }&{ACC AAC GTG TAG AAG GCA TCC}\\
\cline{2-4}
&{NM\_005343}&{AGG ATG CCT TCT ACA CGT TGG}&{CAT GTC CTG AGC TTG TGC CT}\\
\hline	
\multirow{2}{*}{NSD-1}&{NM\_022455}&{TCG CCA TTC TTG CCA TTA GC}&{TTT TCA TTG CTG CCG TCC AC}\\
\cline{2-4}
&{NM\_172349}&{ATT GTC TGC TGC CCT TTT CC}&{TGG AAT CTG GAT CAT CCC GA}\\
\hline	
\multirow{2}{*}{TSC2$^{*}$}&{NM\_000548}&{CTC TCC ACC CGT GAA AGA ATT C}&{GAC CAC ATG TTC AGA CAC ACT G}\\
\cline{2-4}
&{NM\_001077183}&{AAC GAG AGA CCC AAG AGG AT}&{GA CGT ATC GAG CCA TCA TGT C}\\
\hline	
\multirow{2}{*}{WHSC1L1}&{NM\_023034}&{ATG TAA AAC TGG GGC AGC AC}&{AAG CAC CAA CAG AAC AAC GC}\\
\cline{2-4}
&{NM\_017778}&{TTT CGG TTT GAG CTG GAT GG}&{TTT GGG CTG TTT GGC AAA CC}\\
\hline											
\end{tabular}
\end{table}

\noindent{\bf S2 Table. Primer sets of the transcripts in five genes of OVCAR8 cancer cell line.} * Gene contains more transcript(s) which can not be quantified by qRT-PCR.
\vspace{4mm}
\begin{table}[H]
\tiny
\centering
\begin{tabular}{|c|c|c|c|}
\hline
{Gene Name}&{Transcript Name}&{Primer Sequence - Forward}&{Primer Sequence - Reverse}\\
\hline
\multirow{2}{*}{ERBB2}&{NM\_001005862}&{5-CACAGATAAAACGGGGGCAC}&{5-CAGGGTCTGAGTCTCTGTGCT}\\
&{NM\_004448}&{5-GAGGGCTGCTTGAGGAAGTAT}&{5-TTTCTCCGGTCCCAATGGAG}\\
\hline
\multirow{2}{*}{NSD1}&{NM\_022455}&{5-GACACGGTGCAGTCAAATCG}&{5-GCTGCCGTCCACTTCATTTC}\\
&{NM\_172349}&{5-AGAAGAAATTGTCTGCTGCCC}&{5-GGATCATCCGAAAGGGCTGT}\\
\hline
\multirow{2}{*}{U2AF1$^{*}$}&{NM\_001025203}&{5-TTGGAGCATGTCGTCATGGAG}&{5-CTGTGCACTGTTTTGGGGATT}\\
&{NM\_006758}&{5-TGCCCTCTTGAACATTTACCGT}&{5-CTGCATCTCCACATCGCTCA}\\
\hline
\multirow{2}{*}{PDGFB}&{NM\_002608}&{5-CTCCGCGCTTTCCGATTTTG}&{5-AGAGGAAAAGGAACACGGCA}\\
&{NM\_033016}&{5-GACTGAGCAGGAATGGTGAGAT}&{5-TCAAAGGAGCGGATCGAGTG}\\
\hline
\multirow{2}{*}{DNMT3A$^{*}$}&{NM\_153759}&{5-GCAGCTACTTCCAGAGCTTCA}&{5-TTTCAGGCTACGATCCACGC}\\
&{NM\_175630}&{5-GGGCAGCAGATACCCTGTTT}&{5-GGCTGGGCAGTACACAGAAT}\\
\hline
\multirow{2}{*}{GNAS$^{*}$}&{NM\_016592}&{5-CGAGTCTTAGGCTGCGGAAT}&{5-GCACCTACCTTCCTGACCAC}\\
&{NM\_080425}&{5-CACTCCCGTCAACATGGACA}&{5-GTACCCCGGAGAGGGTACTT}\\
\hline
\multirow{2}{*}{RBM15}&{NM\_001201545}&{5-ATGCCTTCCCACCTTGTGAG}&{5-TCAACCAGTTTTGCACGGAC}\\
&{NM\_022768}&{5-AACAAGAAGAGAGAAAACTTGGCG}&{5-TTTCCTCCCTTTAGGGACACC}\\
\hline
\multirow{2}{*}{RET}&{NM\_020630}&{5-TGCCCAGCAACTTAGGATGG}&{5-TTGATTCCCACCCCAGAAGC}\\
&{NM\_020975}&{5-AATGGAAAGTCTACCGGCCC}&{5-CAGAGCTCTTACCCGGTGTG}\\
\hline
\multirow{2}{*}{TCF3}&{NM\_001136139}&{5-GAGAAAGACCTGAGGGACCG}&{5-GGCCTCGTTAATATCCCGCA}\\
&{NM\_003200}&{5-CAACTGCACCTCAACAGCGA}&{5-CTCCAAGTTCAGGATGACCGA}\\
\hline
\multirow{2}{*}{WHSC1L1}&{NM\_017778}&{5-GCCTCTCAGTACAGCACTCC}&{5-GCCTGCCCATGTTAATGCTG}\\
&{NM\_023034}&{5-AGAAAGGTGCCAGCGAGATT}&{5-GCAGGTCACTCAGTCCTCTA}\\
\hline
\multirow{2}{*}{CBFB}&{NM\_001755}&{5-GGATGCATTAGCACAACAGGC}&{5-GCCAGCAGCTGTGAAACTCT}\\
&{NM\_022845}&{5-CGGGAGGAAATGGAGGCAAG}&{5-GTAAAGATGGGCAGCACACAT}\\
\hline
\multirow{2}{*}{TP53}&{Iso Group1}&{5-GATGAAGCTCCCAGAATGCC}&{5-GTAGCTGCCCTGGTAGGTTT}\\
&{Iso Group2}&{5-GAGGTGTAGACGCCAACTCT}&{5-AAGTCAGGGCACAAGTGAACA}\\
\hline
\multirow{2}{*}{NF1$^{*}$}&{NM\_000267}&{5-TGAGGAAAACCAGCGGAACC}&{5-GCTGGCTAACCACCTGGTATAAA}\\
&{NM\_001128147}&{5-GTGGAATCCTGATGCTCCTGT}&{5-AAAACCATAAAACCTTTGGAAGTGT}\\
\hline											
\end{tabular}
\end{table}
\noindent{\bf S3 Table. Primer sets of the transcripts in thirteen genes of MCF7 cancer cell line.} * Gene contains more transcript(s) which can not be quantified by qRT-PCR.
\begin{table}[H]
\tiny
\centering
\begin{tabular}{|c|c|c|c|c|}
\hline
{\bf KEGG Pathway}&{\bf \# of gene in KEGG}&{\bf \# of overlapped gene}&{\bf Interactions}&{\bf Density}\\
\hline
{Long-term potentiation}&{67}&{30}&{552}&{61.33\%}\\
{Fc epsilon RI signaling pathway}&{70}&{51}&{1426}&{54.83\%}\\
{Fc gamma R-mediated phagocytosis}&{91}&{45}&{1002}&{49.48\%}\\
{Axon guidance}&{127}&{56}&{1526}&{48.66\%}\\
{VEGF signaling pathway}&{61}&{39}&{736}&{48.39\%}\\
{GnRH signaling pathway}&{92}&{53}&{1356}&{48.27\%}\\
{Phosphatidylinositol signaling system}&{82}&{26}&{318}&{47.04\%}\\
{ErbB signaling pathway}&{87}&{70}&{2272}&{46.37\%}\\
{Inositol phosphate metabolism}&{63}&{14}&{86}&{43.88\%}\\
{Neuroactive ligand-receptor interaction}&{275}&{38}&{626}&{43.35\%}\\
{Glycerophospholipid metabolism}&{95}&{12}&{62}&{43.06\%}\\
{Olfactory transduction}&{407}&{15}&{92}&{40.89\%}\\
{Neurotrophin signaling pathway}&{120}&{88}&{3128}&{40.39\%}\\
{Central carbon metabolism in cancer}&{67}&{38}&{580}&{40.17\%}\\
{Natural killer cell mediated cytotoxicity}&{134}&{47}&{848}&{38.39\%}\\
{Glutamatergic synapse}&{116}&{21}&{168}&{38.10\%}\\
{Retrograde endocannabinoid signaling}&{103}&{25}&{238}&{38.08\%}\\
{RNA polymerase}&{32}&{11}&{46}&{38.02\%}\\
{Glioma}&{65}&{46}&{800}&{37.81\%}\\
{Ras signaling pathway}&{228}&{119}&{5182}&{36.59\%}\\
{Inflammatory mediator regulation of TRP channels}&{99}&{45}&{722}&{35.65\%}\\
{Long-term depression}&{60}&{25}&{220}&{35.20\%}\\
{Gap junction}&{89}&{42}&{620}&{35.15\%}\\
{B cell receptor signaling pathway}&{72}&{45}&{706}&{34.86\%}\\
{Tight junction}&{138}&{49}&{830}&{34.57\%}\\
{Glycerolipid metabolism}&{59}&{13}&{58}&{34.32\%}\\
{mTOR signaling pathway}&{60}&{35}&{418}&{34.12\%}\\
{T cell receptor signaling pathway}&{104}&{80}&{2178}&{34.03\%}\\
{Vascular smooth muscle contraction}&{121}&{43}&{622}&{33.64\%}\\
{Non-small cell lung cancer}&{56}&{45}&{662}&{32.69\%}\\
{Chemokine signaling pathway}&{189}&{86}&{2408}&{32.56\%}\\
{Cell adhesion molecules (CAMs)}&{142}&{32}&{332}&{32.42\%}\\
{Choline metabolism in cancer}&{101}&{56}&{1012}&{32.27\%}\\
{MAPK signaling pathway}&{257}&{156}&{7830}&{32.17\%}\\
{Epithelial cell signaling in Helicobacter pylori infection}&{68}&{36}&{414}&{31.94\%}\\
{Oxytocin signaling pathway}&{159}&{63}&{1264}&{31.85\%}\\
{Type II diabetes mellitus}&{48}&{23}&{166}&{31.38\%}\\
{Adherens junction}&{73}&{48}&{722}&{31.34\%}\\
{Lysine degradation}&{52}&{11}&{36}&{29.75\%}\\
{Circadian entrainment}&{97}&{28}&{232}&{29.59\%}\\
{Prolactin signaling pathway}&{72}&{52}&{792}&{29.29\%}\\
{Insulin signaling pathway}&{140}&{75}&{1634}&{29.05\%}\\
{Cholinergic synapse}&{113}&{42}&{512}&{29.02\%}\\
{Morphine addiction}&{93}&{21}&{126}&{28.57\%}\\
{Bladder cancer}&{38}&{34}&{322}&{27.85\%}\\
{Platelet activation}&{131}&{57}&{892}&{27.45\%}\\
{Progesterone-mediated oocyte maturation}&{88}&{63}&{1074}&{27.06\%}\\
{Toll-like receptor signaling pathway}&{106}&{56}&{842}&{26.85\%}\\
{Rap1 signaling pathway}&{211}&{119}&{3722}&{26.28\%}\\
{Leukocyte transendothelial migration}&{118}&{51}&{680}&{26.14\%}\\
{Pancreatic cancer}&{66}&{58}&{870}&{25.86\%}\\
{Complement and coagulation cascades}&{69}&{23}&{136}&{25.71\%}\\
{Acute myeloid leukemia}&{57}&{48}&{586}&{25.43\%}\\
{Renal cell carcinoma}&{66}&{55}&{754}&{24.93\%}\\
{Endometrial cancer}&{52}&{41}&{414}&{24.63\%}\\
{Sphingolipid signaling pathway}&{120}&{66}&{1068}&{24.52\%}\\
{Focal adhesion}&{207}&{103}&{2600}&{24.51\%}\\
{Regulation of actin cytoskeleton}&{215}&{95}&{2210}&{24.49\%}\\
{FoxO signaling pathway}&{134}&{94}&{2102}&{23.79\%}\\
{Amyotrophic lateral sclerosis (ALS)}&{51}&{20}&{94}&{23.50\%}\\
{Osteoclast differentiation}&{131}&{82}&{1544}&{22.96\%}\\
{Melanoma}&{71}&{50}&{566}&{22.64\%}\\
{Serotonergic synapse}&{114}&{29}&{190}&{22.59\%}\\
{Gastric acid secretion}&{75}&{24}&{130}&{22.57\%}\\
{Hepatitis C}&{133}&{73}&{1194}&{22.41\%}\\
{Dopaminergic synapse}&{131}&{44}&{426}&{22.00\%}\\
{Calcium signaling pathway}&{180}&{51}&{570}&{21.91\%}\\
{Pathogenic Escherichia coli infection}&{55}&{22}&{104}&{21.49\%}\\
{Influenza A}&{175}&{76}&{1172}&{20.29\%}\\
{Estrogen signaling pathway}&{100}&{51}&{524}&{20.15\%}\\
{Chronic myeloid leukemia}&{73}&{63}&{796}&{20.06\%}\\
{Aldosterone-regulated sodium reabsorption}&{39}&{22}&{96}&{19.83\%}\\
{Ovarian steroidogenesis}&{51}&{18}&{64}&{19.75\%}\\
{NOD-like receptor signaling pathway}&{57}&{31}&{186}&{19.35\%}\\
{cAMP signaling pathway}&{200}&{82}&{1298}&{19.30\%}\\
{Colorectal cancer}&{62}&{52}&{498}&{18.42\%}\\
{Proteoglycans in cancer}&{204}&{130}&{3082}&{18.24\%}\\
{Prostate cancer}&{89}&{70}&{884}&{18.04\%}\\
{RIG-I-like receptor signaling pathway}&{70}&{35}&{220}&{17.96\%}\\
{Signaling pathways regulating pluripotency of stem cells}&{142}&{84}&{1250}&{17.72\%}\\
{cGMP-PKG signaling pathway}&{167}&{56}&{542}&{17.28\%}\\
{Thyroid hormone signaling pathway}&{119}&{62}&{662}&{17.22\%}\\
{TGF-beta signaling pathway}&{80}&{51}&{446}&{17.15\%}\\
{Endocytosis}&{213}&{84}&{1204}&{17.06\%}\\
{ECM-receptor interaction}&{87}&{23}&{90}&{17.01\%}\\
{Adipocytokine signaling pathway}&{70}&{38}&{242}&{16.76\%}\\
{Hepatitis B}&{146}&{98}&{1510}&{15.72\%}\\
{Ubiquitin mediated proteolysis}&{137}&{66}&{654}&{15.01\%}\\
{MicroRNAs in cancer}&{297}&{109}&{1484}&{12.49\%}\\
{Pathways in cancer}&{398}&{260}&{5702}&{8.43\%}\\
\hline
\end{tabular}
\end{table}
\noindent{\bf S4 Table. Overlapped KEGG pathways with large transcript network.} We consider the subnetwork of genes that are members of one KEGG pathway and calculated the density of DDIs in the subnetwork.
\vspace{4mm}
\begin{table}[H]
\scriptsize
\centering
\begin{tabular}{|c|c|c|c|c|c|c|}
\hline
\multirow{2}{*}{\bf Gene Name}&\multirow{2}{*}{\bf Transcript Name}&\multicolumn{4}{c|}{\bf Estimated Proportion}&{\bf qRT-PCR}\\
\cline{3-6}
&&{\bf Net-RSTQ}&{\bf base EM}&{\bf Cufflinks}&{\bf RSEM}&{\bf Results}\\
\hline
\multirow{2}{*}{ABL1}&{NM\_007313}&{64.46$\%$}&{94.45$\%$}&{16.48$\%$}&{53.14$\%$}&{56$\pm$4.4$\%$}\\
&{NM\_005157}&{35.54$\%$}&{5.55$\%$}&{83.52$\%$}&{46.86$\%$}&{44$\pm$4.4$\%$}\\
\hline
\multirow{2}{*}{CBLC}&{NM\_012116}&{73.12$\%$}&{93.23$\%$}&{87.59$\%$}&{87.54$\%$}&{51$\pm$9.8$\%$}\\
&{NM\_001130852}&{26.88$\%$}&{6.77$\%$}&{12.41$\%$}&{12.46$\%$}&{49$\pm$9.8$\%$}\\
\hline
\multirow{2}{*}{KDM5C}&{NM\_004187}&{80.52$\%$}&{99.22$\%$}&{99.95$\%$}&{91.52$\%$}&{86$\pm$4.5$\%$}\\
&{NM\_001146702}&{19.48$\%$}&{0.78$\%$}&{0.05$\%$}&{8.48$\%$}&{14$\pm$4.5$\%$}\\
\hline
\multirow{2}{*}{KRAS}&{NM\_033360}&{51.36$\%$}&{80.23$\%$}&{58.82$\%$}&{19.67$\%$}&{36$\pm$4.2$\%$}\\
&{NM\_004985}&{48.64$\%$}&{19.77$\%$}&{41.18$\%$}&{80.33$\%$}&{64$\pm$4.2$\%$}\\
\hline
\multirow{3}{*}{NPM1}&{NM\_002520 (Iso1)}&{34.92$\%$}&{55.84$\%$}&{0$\%$}&{84.62$\%$}&{52$\%$$^{*}$}\\
&{NM\_199185 (Iso2)}&{29.09$\%$}&{6.56$\%$}&{53.97$\%$}&{1.52$\%$}&{45$\%$}\\
&{NM\_001037738 (Iso3)}&{35.99$\%$}&{37.60$\%$}&{46.03$\%$}&{13.86$\%$}&{3.2$\%$}\\
\hline
\multirow{2}{*}{TCF3}&{NM\_003200}&{78.31$\%$}&{98.11$\%$}&{96.42$\%$}&{90.06$\%$}&{56$\pm$6.5$\%$}\\
&{NM\_001136139}&{21.69$\%$}&{1.89$\%$}&{3.58$\%$}&{9.94$\%$}&{44$\pm$6.5$\%$}\\
\hline
\multirow{2}{*}{WHSC1L1}&{NM\_023034}&{39.98$\%$}&{73.61$\%$}&{4.50$\%$}&{37.72$\%$}&{46$\pm$6.1$\%$}\\
&{NM\_017778}&{60.02$\%$}&{26.39$\%$}&{95.50$\%$}&{62.28$\%$}&{54$\pm$6.1$\%$}\\
\hline
\end{tabular}
\end{table}

\noindent{\bf S5 Table. qRT-PCR results on H9 stem cell line.} * Standard deviation of Iso1+Iso3 is 5.7\% and Iso3 is 4.4\%
\vspace{4mm}
\begin{table}[H]
\scriptsize
\centering
\begin{tabular}{|c|c|c|c|c|c|c|}
\hline
\multirow{2}{*}{\bf Gene Name}&\multirow{2}{*}{\bf Transcript Name}&\multicolumn{4}{c|}{\bf Estimated Proportion}&{\bf qRT-PCR}\\
\cline{3-6}
&&{\bf Net-RSTQ}&{\bf base EM}&{\bf Cufflinks}&{\bf RSEM}&{\bf Results}\\
\hline
\multirow{2}{*}{HNRNPA2B1}&{NM\_031243}&{60.07$\%$}&{81.42$\%$}&{0$\%$}&{42.37$\%$}&{82.83$\%$}\\
&{NM\_002137}&{39.93$\%$}&{18.58$\%$}&{100$\%$}&{57.63$\%$}&{17.17$\%$}\\
\hline
\multirow{2}{*}{HRAS$^{*}$}&{NM\_176795}&{54.92$\%$}&{100$\%$}&{11.62$\%$}&{11.72$\%$}&{50.16$\%$}\\
&{NM\_005343}&{45.08$\%$}&{0$\%$}&{88.38$\%$}&{88.28$\%$}&{49.84$\%$}\\
\hline
\multirow{2}{*}{NSD1}&{NM\_022455}&{55.40$\%$}&{96.39$\%$}&{99.94$\%$}&{61.65$\%$}&{99.98$\%$}\\
&{NM\_172349}&{44.60$\%$}&{3.61$\%$}&{0.06$\%$}&{38.35$\%$}&{0.02$\%$}\\
\hline
\multirow{2}{*}{TSC2$^{*}$}&{NM\_000548}&{52.39$\%$}&{92.36$\%$}&{0.01$\%$}&{4.62$\%$}&{45.36$\%$}\\
&{NM\_001077183}&{47.61$\%$}&{7.64$\%$}&{99.99$\%$}&{95.38$\%$}&{54.64$\%$}\\
\hline
\multirow{2}{*}{WHSC1L1}&{NM\_023034}&{48.84$\%$}&{63.85$\%$}&{0.01$\%$}&{35.69$\%$}&{18.81$\%$}\\
&{NM\_017778}&{51.16$\%$}&{36.15$\%$}&{99.99$\%$}&{64.31$\%$}&{81.19$\%$}\\
\hline											
\end{tabular}
\end{table}
\noindent{\bf S6 Table. qRT-PCR results on OVCAR8 cancer cell line.} * Gene contains more transcript which can not be quantified by qRT-PCR.
\vspace{4mm}
\begin{table}[H]
\scriptsize
\centering
\begin{tabular}{|c|c|c|c|c|c|c|}
\hline
\multirow{2}{*}{\bf Gene Name}&\multirow{2}{*}{\bf Transcript Name}&\multicolumn{4}{c|}{\bf Estimated Proportion}&{\bf qRT-PCR}\\
\cline{3-6}
&&{\bf Net-RSTQ}&{\bf base EM}&{\bf Cufflinks}&{\bf RSEM}&{\bf Results}\\
\hline
\multirow{2}{*}{ERBB2}&{NM\_001005862}&{26.16$\%$}&{7.79$\%$}&{3.35$\%$}&{6.34$\%$}&{7.35$\pm$0.75$\%$}\\
&{NM\_004448}&{73.84$\%$}&{92.21$\%$}&{96.65$\%$}&{93.66$\%$}&{92.65$\pm$6.0$\%$}\\
\hline
\multirow{2}{*}{NSD1}&{NM\_022455}&{46.77$\%$}&{20.08$\%$}&{21.94$\%$}&{17.14$\%$}&{58.34$\pm$0.70$\%$}\\
&{NM\_172349}&{53.23$\%$}&{79.92$\%$}&{78.06$\%$}&{82.86$\%$}&{41.66$\pm$0.75$\%$}\\
\hline
\multirow{2}{*}{U2AF1$^{*}$}&{NM\_001025203}&{26.68$\%$}&{21.11$\%$}&{73.15$\%$}&{25.33$\%$}&{39.13$\pm$0.50$\%$}\\
&{NM\_006758}&{73.32$\%$}&{78.89$\%$}&{26.85$\%$}&{74.67$\%$}&{60.87$\pm$2.5$\%$}\\
\hline
\multirow{2}{*}{PDGFB}&{NM\_002608}&{21.51$\%$}&{18.03$\%$}&{43.93$\%$}&{18.12$\%$}&{97.69$\pm$6.5$\%$}\\
&{NM\_033016}&{78.49$\%$}&{81.97$\%$}&{56.07$\%$}&{81.88$\%$}&{2.31$\pm$1.4$\%$}\\
\hline
\multirow{2}{*}{DNMT3A$^{*}$}&{NM\_153759}&{99.53$\%$}&{99.77$\%$}&{99.45$\%$}&{99.46$\%$}&{11.06$\pm$3.0$\%$}\\
&{NM\_175630}&{0.47$\%$}&{0.23$\%$}&{0.55$\%$}&{0.54$\%$}&{88.94$\pm$4.0$\%$}\\
\hline
\multirow{2}{*}{GNAS$^{*}$}&{NM\_016592}&{69.65$\%$}&{87.11$\%$}&{93.62$\%$}&{89.66$\%$}&{98.27$\pm$5.0$\%$}\\
&{NM\_080425}&{30.35$\%$}&{12.89$\%$}&{6.38$\%$}&{10.34$\%$}&{1.73$\pm$0$\%$}\\
\hline
\multirow{2}{*}{RBM15}&{NM\_001201545}&{51.65$\%$}&{77.37$\%$}&{63.21$\%$}&{73.85$\%$}&{30.21$\pm$0.55$\%$}\\
&{NM\_022768}&{48.35$\%$}&{22.63$\%$}&{36.79$\%$}&{26.15$\%$}&{69.79$\pm$1.8$\%$}\\
\hline
\multirow{2}{*}{RET}&{NM\_020630}&{57.13$\%$}&{60.36$\%$}&{70.78$\%$}&{67.76$\%$}&{37.04$\pm$1.3$\%$}\\
&{NM\_020975}&{42.87$\%$}&{39.64$\%$}&{29.22$\%$}&{32.24$\%$}&{62.96$\pm$2.3$\%$}\\
\hline
\multirow{2}{*}{TCF3}&{NM\_001136139}&{35.90$\%$}&{31.51$\%$}&{6.70$\%$}&{36.31$\%$}&{26.85$\pm$1.1$\%$}\\
&{NM\_003200}&{64.10$\%$}&{68.49$\%$}&{93.30$\%$}&{63.69$\%$}&{73.15$\pm$4.0$\%$}\\
\hline
\multirow{2}{*}{WHSC1L1}&{NM\_017778}&{59.60$\%$}&{54.38$\%$}&{77.61$\%$}&{62.43$\%$}&{77.36$\pm$2.0$\%$}\\
&{NM\_023034}&{40.40$\%$}&{45.62$\%$}&{22.39$\%$}&{37.57$\%$}&{22.64$\pm$0.18$\%$}\\
\hline
\multirow{2}{*}{CBFB}&{NM\_001755}&{51.89$\%$}&{52.80$\%$}&{13.84$\%$}&{57.67$\%$}&{63.75$\pm$1.1$\%$}\\
&{NM\_022845}&{48.11$\%$}&{47.20$\%$}&{86.16$\%$}&{42.33$\%$}&{36.25$\pm$1.2$\%$}\\
\hline
\multirow{2}{*}{TP53}&{Iso Group1}&{96.90$\%$}&{99.37$\%$}&{99.10$\%$}&{97.41$\%$}&{98.12$\pm$6.0$\%$}\\
&{Iso Group2}&{3.10$\%$}&{0.63$\%$}&{0.90$\%$}&{2.59$\%$}&{1.88$\pm$0.30$\%$}\\
\hline
\multirow{2}{*}{NF1$^{*}$}&{NM\_000267}&{98.08$\%$}&{98.56$\%$}&{15.14$\%$}&{94.55$\%$}&{85.76$\pm$2.5$\%$}\\
&{NM\_001128147}&{1.92$\%$}&{1.44$\%$}&{84.86$\%$}&{5.45$\%$}&{14.24$\pm$2.0$\%$}\\
\hline											
\end{tabular}
\end{table}
\noindent{\bf S7 Table. qRT-PCR results on MCF7 cancer cell line.} * Gene contains more transcript(s) which can not be quantified by qRT-PCR.
\vspace{4mm}
\begin{table}[H]
\centering
\begin{tabular}{|c|c|c|c|c|c|}
\hline
&\multicolumn{5}{c|}{\bf Regularized framework}\\
\hline
\multirow{17}{*}{\bf Net-RSTQ}&{\bf $\lambda$}&{\bf 0.1}&{\bf 1}&{\bf 10}&{\bf 100}\\
\cline{2-6}
&{\bf 1e-04}&{0.9689}&{0.9536}&{0.9442}&{0.8668}\\
\cline{2-6}
&{\bf 1e-03}&{0.9696}&{0.9549}&{0.9453}&{0.8680}\\
\cline{2-6}
&{\bf 5e-03}&{0.9722}&{0.9583}&{0.9488}&{0.8715}\\
\cline{2-6}
&{\bf 0.01}&{0.9751}&{0.9620}&{0.9526}&{0.8751}\\
\cline{2-6}
&{\bf 0.05}&{0.9887}&{0.9804}&{0.9720}&{0.8939}\\
\cline{2-6}
&{\bf 0.1}&{0.9943}&{0.9895}&{0.9819}&{0.9038}\\
\cline{2-6}
&{\bf 0.2}&{\bf 0.9959}&{0.9947}&{0.9882}&{0.9111}\\
\cline{2-6}
&{\bf 0.3}&{0.9950}&{\bf 0.9954}&{0.9899}&{0.9141}\\
\cline{2-6}
&{\bf 0.4}&{0.9938}&{0.9951}&{0.9905}&{0.9164}\\
\cline{2-6}
&{\bf 0.5}&{0.9925}&{0.9945}&{0.9909}&{0.9188}\\
\cline{2-6}
&{\bf 0.6}&{0.9910}&{0.9935}&{0.9912}&{0.9218}\\
\cline{2-6}
&{\bf 0.7}&{0.9889}&{0.9920}&{\bf 0.9914}&{0.9258}\\
\cline{2-6}
&{\bf 0.8}&{0.9854}&{0.9891}&{0.9911}&{0.9319}\\
\cline{2-6}
&{\bf 0.9}&{0.9770}&{0.9815}&{0.9883}&{0.9424}\\
\cline{2-6}
&{\bf 0.99}&{0.9328}&{0.9387}&{0.9543}&{\bf 0.9601}\\
\cline{2-6}
&{\bf 0.999}&{0.9121}&{0.9181}&{0.9349}&{0.9578}\\
\hline
\end{tabular}
\end{table}
\noindent{\bf S8 Table. Correlation Coefficients between the results of Net-RSTQ and the alternative regularized framework with different $\lambda$s.} The highest correlation coefficients for each $\lambda$ in the alternative regularized framework is bold.
\end{document}